\documentclass[aip,graphicx]{revtex4-2}

\draft 

\usepackage{comment}

\usepackage[utf8]{inputenc} 
\usepackage{hyperref}       
\usepackage{url}            
\usepackage{booktabs}       
\usepackage{amsfonts}       
\usepackage{nicefrac}       
\usepackage{microtype}      
\usepackage{xcolor}         

\usepackage{amsmath}       
\usepackage{amsthm}       
 
\usepackage{bbold}

\usepackage[ruled,vlined]{algorithm2e}

\newtheorem{lemma}{Lemma}
\newtheorem{corollary}{Corollary}

\usepackage{soul}

\newcommand{\embeddingDimension}{\ell}
\newcommand{\R}{\mathbb{R}}

\usepackage{graphicx}

\newcommand{\change}{} 
\newcommand{\review}[1]{{#1}} 

\newcommand*\diff{{\change \mathop{}\!\mathrm{d}}}
\newcommand{\diffdiscrete}{{\change \delta}}
\newcommand{\deltat}{{\change h}}
\newcommand{\ndim}{{\change n}}
\newcommand{\idx}{{\change k}}

\newcommand{\numpoints}{{\change N}}
\newcommand{\numparticles}{{\change N}}

\newcommand{\migrationrate}{{\change D}}
\newcommand{\SIRvar}{y}

\newcommand{\expectation}[1]{{\change \mathbb{E}\left[#1\right]}}

\newcommand{\X}{{\change {x}}}
\newcommand{\diffx}{{\change \diff x}}

\newcommand{\trainingdata}{{\change D}}

\newcommand{\weights}{{\change \theta}}
\newcommand{\drift}{{\change f}}
\newcommand{\std}{{\change \sigma}}
\newcommand{\driftNN}{{\change f_\weights}}
\newcommand{\stdNN}{{\change \sigma_\weights}}

\newcommand{\lossfunction}{{\change \mathcal{L}}}

\newcommand{\normal}{{\change \mathcal{N}}}

\newcommand{\eqnref}[1]{{\change Eq.~\eqref{#1}}}
\newcommand{\secref}[1]{{\change Sec.~\ref{#1}}}
\newcommand{\figref}[1]{{\change Fig.~\ref{#1}}}

\begin{document}

\title{Learning effective stochastic differential equations from microscopic simulations: linking stochastic numerics to deep learning}

\author{Felix Dietrich}
\email[]{felix.dietrich@tum.de}
\thanks{Source code at \url{https://gitlab.com/felix.dietrich/sde-identification}.}
\affiliation{Technical University of Munich, Munich, Germany}

\author{Alexei Makeev}
\email[]{amak@cs.msu.ru}
\affiliation{Moscow State University, Moscow, Russia}

\author{George Kevrekidis}
\email[]{georgek@math.umass.edu}
\affiliation{University of Massachusetts, Amherst, MA, USA}

\author{Nikolaos Evangelou}
\email[]{nevange2@jh.edu}
\affiliation{Johns Hopkins University, Baltimore, MD, USA}

\author{Tom Bertalan}
\email[]{tom@tombertalan.com}
\affiliation{Johns Hopkins University, Baltimore, MD, USA}

\author{Sebastian Reich}
\email[]{sereich@uni-potsdam.de}
\affiliation{University of Potsdam, Potsdam, Germany}

\author{Ioannis G.~Kevrekidis}
\email[]{yannisk@jhu.edu}
\affiliation{Johns Hopkins University, Baltimore, MD, USA}

\date{\today}

\begin{abstract}
We identify effective stochastic 
differential equations (SDE) for coarse observables of fine-grained particle- or agent-based simulations; these SDE then provide useful coarse surrogate models of the fine scale dynamics.
We approximate the drift and diffusivity functions in these effective SDE through neural networks, which
can be thought of as effective stochastic ResNets.
The loss function is inspired by, and embodies, the structure of established stochastic numerical integrators (here, Euler-Maruyama and Milstein); our approximations can thus benefit from backward error analysis of these underlying numerical schemes.
They also lend themselves naturally to ``physics-informed'' gray-box identification when approximate coarse models, such as mean field equations, are available.
\review{Existing numerical integration schemes for Langevin-type equations and for stochastic partial differential equations (SPDE) can also be used for training; we demonstrate this on a stochastically forced oscillator and the stochastic wave equation.}
Our approach does not require long trajectories, works on scattered snapshot data, and is designed to naturally handle different time steps per snapshot.
We consider both the case where the coarse collective observables are known in advance, as well as the case where they must be found in a data-driven manner.
\end{abstract}


\maketitle 

\section{Introduction}\label{sec:introduction}

Residual networks (ResNets~\cite{he-2016}, but see also~\cite{Rico-Martinez-1992}) successfully employ a forward-Euler integrator based approach to create useful deep architectures. This has inspired followup work to include other integrator schemes for deterministic, ordinary differential equations (ODE) such as symplectic integrators~\cite{bertalan-2019,Greydanus-2019,zhu-2020}.
\review{In this work, we propose to include stochastic integrators in the search for appropriate loss functions for training network architectures, to improve system identification and prediction tasks.}

Identification of dynamical systems from data is an established systems task~\cite{Rico-Martinez-1995,mezic-2005,rahimi-2006,bongard-2007,brunton-2016c,klus-2020}. Machine learning both in its ``second spring'' in the the 80-90 years, and its current ``third spring'' has made a major impact in the approaches and algorithms for this task.
\review{From ResNets~\cite{he-2016} and Neural ODEs~\cite{chen-2018,Rico-Martinez-1992} to transformers~\cite{geneva-2022}, ``neural PDEs''~\cite{gonzalez-1998}, and DeepONets~\cite{Lu-2021}}, the identification of dynamical systems from (discrete) spatio-temporal data is a booming business  because it is now comparatively easy to program and train neural networks. 
Learning SDEs and SPDEs is a natural ``next step'' and there is also an array of excellent contributions to this as we outline in the related work section below.
The time-honored SDE estimation techniques for local drift and diffusivity can now be synthesized in a (more or less global) surrogate model.

Given all these recently available options, our contribution is as follows: On the application side, we were motivated by the identification of coarse, or {\em effective} SDEs: atomistic/stochastic/agent-based physics simulations codes whose macroscopic statistics (e.g. leading moment evolution) have long been modelled as effective SDEs, whether with mathematical proofs or heuristically. This is the reason for part of our title, and for some of the examples: lattice simulations of epidemic models.

There is also, however, a methodological motivation whose roots are in the numerical analysis of integration schemes for SDEs/SPDEs. These established algorithms (starting with Euler-Maruyama~\cite{Higham-2001}) 
iteratively produce discrete time data; it makes sense to ``mathematically inform'' the neural network architecture used for identification with the structure of the numerical integrator one would use on an explicitly available, closed form coarse stochastic model.
This way, our work naturally links with, and can take advantage of, the extensive body of work in what is called ``backward error analysis'' of these integration schemes~\cite{Zygalakis-2011}. 


A third motivation for our work: beyond its links to backward error analysis, the approach naturally lends itself to the what one might call ``physics-informed'' or ``gray-box'' identification algorithms: neural network architectures that seamlessly incorporate approximate or partially known analytical models if these are available. 
We developed such an approach for deterministic ODE identification in the early 90s, devising recurrent Runge-Kutta ResNet architectures for this purpose, and extending them to gray boxes and implicit ResNets~\cite{Rico-Martinez-1992,rico-martinez-1994,gonzalez-1998}.
It is a natural next step to extend these techniques so that they can be applied to the identification of effective stochastic models, allowing the possibility of physics-informed gray boxes, and exploiting mathematical support via numerical backward error analysis.
%
Learned, black-box models for agent-based systems may be used in malicious or questionable ways in technology and society. An example could be careless,  uncontrolled generalization in predicting the behavior of physical or societal systems using learned models.
Using gray-box, physics-informed machine learning methods that are amenable to thorough mathematical analyses and interpretability can help mitigate such risks, and our optimization approach is specifically designed for this type of knowledge integration and analysis.

Our contributions are as follows:
\begin{enumerate}
    \item We introduce a neural network training loss function that allows us to approximate the state dependent drift and diffusivity of an SDE from discrete observation data.
    \item The approach does not require long time series: paired snapshots  $(\X_t,\X_{t+\deltat})$, scattered over the domain of interest, are sufficient. Step sizes $\deltat$ can vary for each snapshot.
    \item The approach works point-wise: No distributions need to be approximated from data, nor do we need to compute distances between distributions.
    \item We formulate explicit loss functions for Euler-Maruyama- and Milstein-type integration schemes using Gaussian noise.
    \item \review{For data sampled from SDE of Langevin-type, where the assumption of normally distributed noise in the state variable does not hold anymore, we show how to reformulate simple numerical schemes to enable accurate approximations of the drift and diffusivity in the acceleration terms.}
    \item \review{We show how a given numerical integration scheme for stochastic PDE can be reformulated to match our loss function, and demonstrate it by learning forcing terms in the stochastic wave equation.}
    \item We demonstrate that we can employ the approach to coarse-grain particle/lattice simulations; in particular, we approximate the drift and diffusivity of a two-dimensional mean field SDE for a Gillespie-based~\cite{Gillespie-1976,Gillespie-2000} kinetic mechanistic model, as well as for the kinetic Monte-Carlo lattice simulation of an epidemiological SIR model leading to those kinetics at the mean field limit. 
\end{enumerate}

\section{Related work}

Loss functions and network architectures informed by numerical integration methods were devised several decades ago, e.g.~\cite{Rico-Martinez-1992,rico-martinez-1994}.
Recently, these ideas have been rediscovered and used in deep architectures such as ResNets~\cite{he-2016}, with symplectic schemes for Hamiltonian~\cite{bertalan-2019,Greydanus-2019,zhu-2020} and Poisson networks~\cite{Jin-2020c}.
Scalable gradient approaches were developed for SDE approximation recently~\cite{li2020scalable}, which allow backpropagation through a long series of SDE iterations. 
If only a single time step for each data point is available, as in our work, this type of scalability is not an immediate training issue.

The theory of SDE is well-developed; We refer to standard books on stochastic processes~\cite{Pavliotis-2014}, stochastic calculus~\cite{karatzas-1998}, and stochastic algorithms~\cite{perez2010stochastic}.
In the broader machine learning community, SDE are used mostly for generative modeling, meaning that the actual drift and diffusivity functions learned from the data are not informative, as long as the generated distribution is ``acceptable''.
A classical example are generative-adversarial networks~\cite{goodfellow-2014,jalal-2017}.
In recent years, learning random ODEs (deterministic ODE with random parameters) has been done using an approach with generative-adversarial networks~\cite{liu-2020}.
Long-Short-Term-Memory networks were used in a stochastic setting for the generation of hand-written text~\cite{Graves-2013}.
The authors of~\cite{Song-2021} present a framework that learns an SDE for training and sampling from score-based models, allowing sample generation.

In a classical dynamical systems context, the SDE itself is the desired outcome of approximation and thus, typically, different methods are necessary.
\review{Parameter estimation techniques for Ornstein-Uhlenbeck processes with constant coefficients are readily available~\cite{ditlevsen-2005}.}
Gaussian Processes have been used to identify SDE from observations~\cite{yildiz-2018}, with a target likelihood based on the Euler-Maruyama scheme (not explicitly mentioned by the authors).
Our setting is very different from identifying ODE from noisy observations, for example through Gaussian Processes~\cite{Yang-2021}.
Generative stochastic modeling of strongly nonlinear flows with non-Gaussian statistics is also possible~\cite{Arbabi-2019}.
\review{Researchers studying Neural SDEs have studied their relation to Wasserstein-GANs (generative adversarial networks with Wasserstein loss)~\cite{kidger-2021,kidger-2021a}, and for long time series~\cite{morrill-2021}}.
\review{Since a pre-print of this manuscript appeared, other pre-prints expanded on it by learning SDEs with other noise types, e.g. L\'{e}vy-type~\cite{fang-2022}.}
\review{Another research direction is to identify the correct parameters of a latent SDE, given partial observations of it~\cite{hasan-2022}. This requires special assumptions on the latent space and the observation functions.}
\review{Propagators for SDE have also been covered in the recent literature, with loss functions for moment functions and distributions~\cite{zhu-2022}. These approaches differ from our work in that they do not learn using point-wise on snapshots, but instead require accurate ensemble averages and corresponding large batch sizes.}
\review{Spectral-Galerkin type SPDE solvers using neural networks have also been developed~\cite{salvi-2021}. Such approaches differ from our work in that they learn the solution to a (possibly unknown) SDE, not the SDE itself.}
Density and ensemble approximation techniques~\cite{Yang-2020b} provide a complementary view to the particle-like approach with SDE.
Stochastic dynamical systems can also be described through the associated Fokker-Planck operator. See, e.g.~\cite{berry-2015b}, for a non-parametric forecasting method using the ``shift-map'', a numerical approximation of the operator projected on the eigenfunctions of the Laplace operator.

%
\section{Mathematical problem setting}
%

We discuss SDEs of the form
\begin{equation}\label{eq:sde}
    \diff \X_t=\drift(\X_t)\diff t+\std(\X_t)\diff W_t,
\end{equation}
where $\drift:\mathbb{R}^\ndim\to\mathbb{R}^\ndim$ and $\std:\mathbb{R}^\ndim\to\mathbb{R}^{\ndim\times \ndim}$ are smooth, possibly nonlinear functions; every $\std(\X)$ is positive and bounded away from zero (and a positive-definite matrix if $\ndim>1$), and $W_t$ a Wiener process such that for $t> s$, $W_t-W_s\sim \normal(0,t-s)$. See \cite{Pavliotis-2014} for an introduction to stochastic processes.
%
%
%
%
We assume we have access to a set of $\numpoints$ snapshots $\trainingdata=\lbrace (\X_1^{(\idx)},\X_0^{(\idx)},\deltat^{(\idx)})\rbrace_{k=1}^\numpoints,$
where  $\X_0^{(\idx)}$ are points scattered in the state space of (\ref{eq:sde}) and the value of $\X_1^{(\idx)}$ results from the evolution of \eqref{eq:sde} under a small time-step $\deltat^{(\idx)}>0$, starting at $\X_0^{(\idx)}$.
The snapshots are samples from a distribution, $\X_0 \sim p_0$, and the transition densities $\X_1 \sim p_1(\cdot|\X_0,h)$ are associated with (\ref{eq:sde}) for a given time-step $h>0$, which in turn is chosen from some distribution $p_h$.
The joint data-generating distribution is therefore given by $p(\X_0,\X_1,h) = p_1(\X_1|\X_1,h)p_0(\X_0) p_h(h)$. Alternatively, the data could be collected along a long trajectory $\{\X_{t_i}\}$ of (\ref{eq:sde}) with sample frequency $h_i>0$, that is, $t_{i+1} = t_i + h_i$. 
The problem is to identify the drift $\drift$ and diffusivity $\std$ through two neural networks $\driftNN:\mathbb{R}^\ndim\to\mathbb{R}^\ndim$ and $\stdNN:\mathbb{R}^\ndim\to\mathbb{R}^{\ndim\times \ndim}$, parameterized by their weights $\weights$, only from the data in $\trainingdata$.  We assume the points in $\trainingdata$ are sampled sufficiently densely in the region of interest.

As we will see, each choice of SDE integrator will give a different training algorithm, with a loss based on the particular discretization scheme; training algorithms based on integrators of different accuracy, integrators with different rates and types of convergence (strong or weak), even integrators based on different stochastic calculus type (Ito vs Stratonovich) fall under this umbrella. We will demonstrate these ideas here through two Ito-calculus based training algorithms: the Euler-Maruyama and the Milstein schemes. 

%
\subsection{Identification of drift and diffusivity with the Euler-Maruyama scheme}\label{sec:eulermaruyama}
%

We now formulate and rationalize the loss term that we use to train the neural networks $\driftNN$ and $\stdNN$.
The Euler-Maruyama scheme is a simple method to integrate~\eqref{eq:sde} over a small time $h>0$:
\begin{equation}\label{eq:em scheme}
    \X_1 = \X_0 + \deltat \drift(\X_0) + \std(\X_0)\,\diffdiscrete W_0,
\end{equation}
where $\deltat >0$ is small and $\diffdiscrete W_0$ is a vector of $\ndim$ random variables, all i.i.d. and normally distributed around zero with variance $\deltat$.
The convergence of~\eqref{eq:em scheme} for $\deltat\to 0$ has been studied at length; We refer to standard literature~\cite{Pavliotis-2014}.
We can use this idea to construct a loss function for training the networks $\stdNN$ and $\driftNN$ simultaneously. We initially restrict the discussion to the case $\ndim = 1$ for simplicity. 
Essentially, conditioned on $\X_0$ and $h$, we can think of $\X_1$ as a point drawn from a multivariate normal distribution
\begin{equation}\label{eq:normal for xkp1}
    \X_1\sim \normal\left(\X_0 + \deltat \drift(\X_0), \deltat\std(\X_0)^2\right).
\end{equation}
In the training data set $\trainingdata$, we only have access to triples $(\X_0^{(\idx)},  \X_1^{(\idx)},h^{(\idx)})$, and not the drift $\drift$ and diffusivity $\std$. 
%
To approximate them, we define the probability density $p_\weights$ of the normal distribution~\eqref{eq:normal for xkp1} and then, given the neural networks $\driftNN$ and $\stdNN$, ask that the log-likelihood of the data $\trainingdata$ under the assumption in equation~\eqref{eq:normal for xkp1} is high:
\begin{equation}\label{eq:log marginal likelihood}
    \weights := \text{arg}\max_{\hat{\weights}} \expectation{\log p_{\hat{\weights}}\left(\X_1|\X_0,h\right) }\approx \text{arg}\max_{\hat{\weights}} \frac{1}{\numpoints}\sum_{\idx=1}^{\numpoints} \log p_{\hat{\weights}}\left(\X^{(\idx)}_{1}\left|\X^{(\idx)}_0,\deltat^{(\idx)}\right.\right).
\end{equation}
We can now formulate the loss function that will be minimized to obtain the neural network weights $\weights$. The logarithm of the well-known probability density function of the normal distribution, together with the mean and variance from \eqref{eq:normal for xkp1}, yields the loss to minimize during training,
\begin{equation}\label{eq:lossfunction EM}
    \lossfunction( \weights | \X_0,\X_1,\deltat) := \frac{\left(\X_{1}-\X_0 - \deltat \driftNN(\X_0)\right)^2}{\deltat \,\stdNN(\X_0)^{2}} + \log \left|\deltat\stdNN \left(\X_0\right)^2 \right|+\log(2\pi).
\end{equation}
This formula can easily be generalized to higher dimensions (see supplement), and we use such generalizations for examples in more than one dimension.
Minimizing $\lossfunction$ in~\eqref{eq:lossfunction EM} over the data $\trainingdata$ implies maximization of the log marginal likelihood~\eqref{eq:log marginal likelihood} with the constant terms removed (as they do not influence the minimization) \cite{Pavliotis-2014}.
Likelihood estimation in combination with the normal distribution is used in many variational and generative approaches~\cite{goodfellow-2014,kingma-2013,li2020scalable,yildiz-2018,Yang-2021,Song-2021}.
Note that here, the step size $\deltat^{(\idx)}$ is defined \textit{per snapshot}, so it is possible that it has different values for every index $\idx$. This will be especially useful in coarse-graining Gillespie simulations, where the time step is determined as part of the scheme.

We note that the specific form (\ref{eq:normal for xkp1}) of the transition probabilities $p_\weights$ induced by the Euler-Maruyama method implies that it will, in general, be impossible to exactly reproduce the data-generation density $p(x_0,x_1,h)$ for any finite step-size $h$. This motivates the application of more refined numerical methods, such as the Milstein scheme considered in section \ref{sec:milstein} below. 
Alternatively, one can think of $p_{\weights}$ as being induced by an SDE (\ref{eq:sde}) with modified drift $f$ and diffusion $\sigma$. Modified equations have been derived for a variety of numerical methods, such as the Euler or the Milstein method~\cite{Zygalakis-2011}. In general, there is no second-order modified equation for the Euler-Maruyama method, while one exists for the Milstein scheme. Modified equation analysis in our setting is typically performed with equal step-sizes $h^{(\idx)}$ for all data points.

%
\subsection{Identification of drift and diffusivity with the Milstein scheme}\label{sec:milstein}
%

\begin{figure}[ht]
\centering
\includegraphics[width=0.6\textwidth]{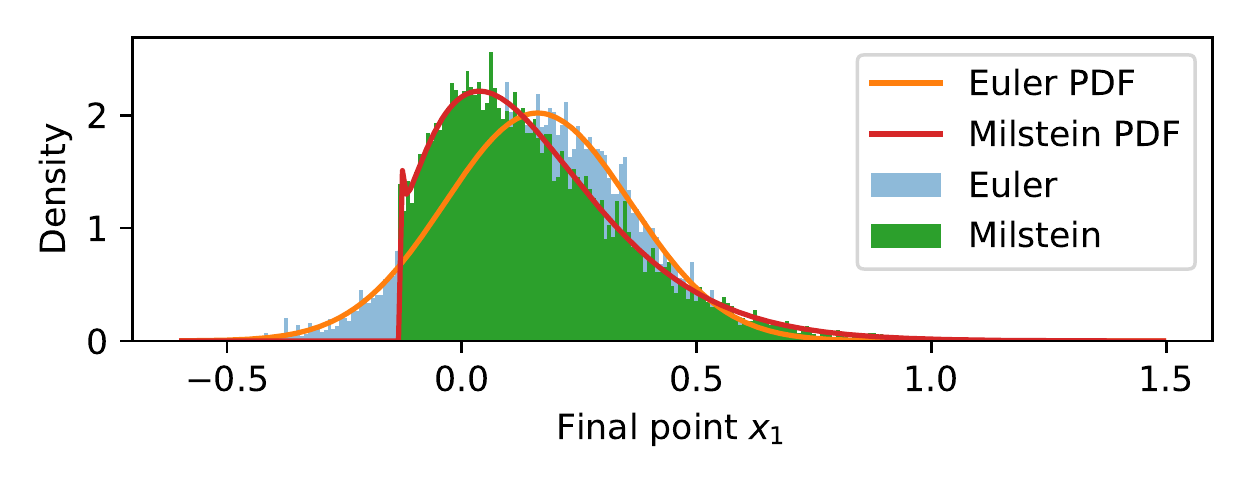}
\vspace*{-.5cm}
\caption{\label{fig:euler vs milstein pdf}Probability density functions and sampled data from the two numerical integrator templates.}
\end{figure}

We now discuss how the loss function can be based on the Milstein integration scheme~\cite{Milshtejn-1975}, to illustrate complexities that arise when deviating from simple Euler schemes, even in one dimension.
For $\ndim=1$, the Milstein scheme is a more accurate method to integrate \eqref{eq:sde}, and is given through
\begin{equation}\label{eq:milstein}
    \X_{1} = \X_0 + \deltat \drift(\X_0) + \std(\X_0)\,\diffdiscrete W_0 + \frac{1}{2}\std(\X_0)\frac{\diff}{\diffx} \std(\X_0)\left((\diffdiscrete W_0)^2-\deltat \right),
\end{equation}
where $\deltat>0$ is small and $\diffdiscrete W_0$ is a random variable, normally distributed around zero with variance $\deltat$.
For the scheme~\eqref{eq:milstein}, it is also possible to write down the distribution of $\X_{1}$, similar to equation~\eqref{eq:normal for xkp1}. It is no longer normal, however, because of the square of the variable $\diffdiscrete W_0$.
To determine the distribution of $\X_1$, we define
\begin{equation*}
A:=\X_0 + \deltat \drift(\X_0) - \deltat  \frac{1}{2}\std(\X_0)\frac{\diff}{\diffx} \std(\X_0);\ 
B:= \sqrt{\deltat} \std(\X_0);\ 
C:= \deltat  \frac{1}{2}\std(\X_0)\frac{\diff}{\diffx} \std(\X_0).
\end{equation*}
Then, for a random variable $z\sim \normal\left(0,1\right)$, we have $\X_{1}=A+Bz+Cz^2.$ Assuming $C\neq 0$, 
\begin{equation*}
   \frac{\X_{\idx+1}-A}{C}=\frac{B}{C}z+z^2=\left(z+\frac{B}{2C}\right)^2-\frac{B^2}{4C^2}\ 
   \iff \frac{\X_{\idx+1}-A+\frac{B^2}{4C}}{C}=\left(z+\frac{B}{2C}\right)^2=:\xi.
\end{equation*}
The variable $\xi$ has a non-central $\chi^2$ distribution with one degree of freedom and non-centrality parameter $\lambda=\left(\frac{B}{2C}\right)^2$. With this derivation, we can see that $\X_{1}$ is distributed according to a linear transformation of a non-central $\chi^2$ distribution. The probability density of $\xi$ is
\begin{equation*}
    p_\xi \left(y;k=1,\lambda=\frac{B^2}{4C^2}\right)=\frac{1}{2}\exp\left(-(y+\lambda)/2\right)\left(\frac{y}{\lambda}\right)^{k/4-1/2} I_{k/2-1}(\sqrt{\lambda y}),
\end{equation*}
where $I_{k/2-1}$ is a modified Bessel function of the first kind.
We can now use the function $p_\xi$ to define the probability density $p_\weights$ of $\X_{1}$, given $\X_0$:
\begin{equation}\label{eq:milstein PDF}
    p_{\text{Milstein},\weights}(\X_1|\X_0,\deltat):=\frac{1}{C} p_\xi\left(\frac{\X_1-A+\frac{B^2}{4C}}{C}; k=1,\lambda=\frac{B^2}{4C^2}\right).
\end{equation}
The corresponding point-wise loss function then is very similar to \eqref{eq:lossfunction EM},
\begin{equation}\label{eq:lossfunction milstein}
    \lossfunction( \weights | \X_0,\X_{1},\deltat) := -\log  p_{\text{Milstein},\weights}(\X_{1}|\X_0,\deltat).
\end{equation}
Equation~\eqref{eq:milstein PDF} defines a probability density that can vanish over large portions of its domain, which leads to issues both with log-marginal likelihood estimation and with gradient computations. These issues can be mitigated through regularization, but also through kernel density approximations, which we outline in \secref{sec:numerical pdf approximation}.
To illustrate the difference of the probability distributions for the Euler-Maruyama and the Milstein templates, we plot them for $\deltat=0.15$, $\drift(\X)=x+1$, $\std(\X)= \frac{1}{4}\X^2+\X+\frac{1}{2}$ in \figref{fig:euler vs milstein pdf} against $10^5$ sampled points using the two templates with the starting point $\X_0=0$.

%
\subsection{Approximation of the probability density of numerical integration schemes}\label{sec:numerical pdf approximation}
%

Inspired by the iteration procedures use to discover kernels for deep Gaussian processes~\cite{lee-2017c} and kernel density estimates~\cite{Higham-2001,schoelkopf-2018}, we now describe an iterative method to approximate the probability density of \review{other numerical integration schemes, such as Runge-Kutta type integrators~\cite{roberts-2012a}}. This will be useful in cases for which the density cannot be derived in analytical, closed form, or for schemes with potentially locally vanishing probability density, like the Milstein method.

\begin{figure}[ht]
    \centering
    \includegraphics[width=.49\textwidth]{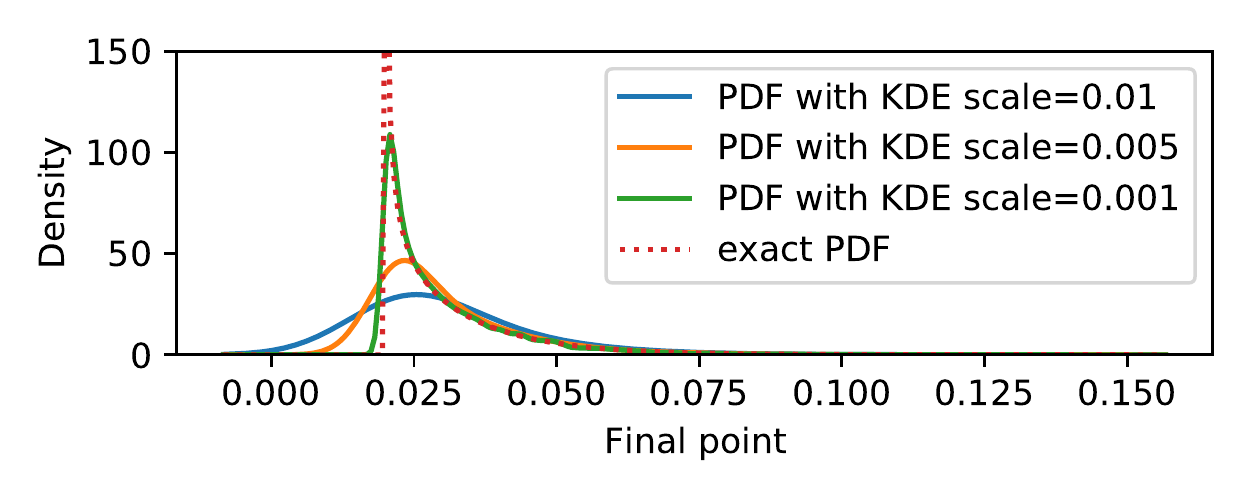}
    \includegraphics[width=.49\textwidth]{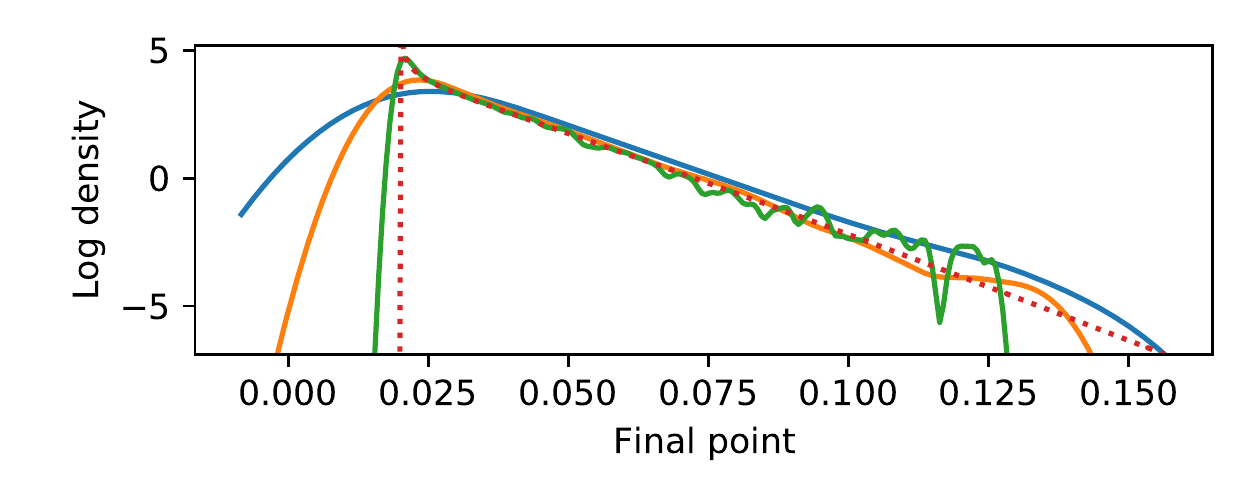}
    \caption{\label{fig:pdf approximation milstein}Approximations of~\eqref{eq:milstein PDF} at varying scales $\sigma_r$, in standard and log-scale.}
\end{figure}

Consider an arbitrary (but convergent and accurate) numerical integration scheme $\phi$ for \eqref{eq:sde}, so that for a fixed step size $\deltat>0$ and a noise process $\diffdiscrete W_0$, a single time-step becomes
\begin{equation} \label{eq:general time stepping}
    \X_{1}=\phi\left(\X_0; \deltat, \diffdiscrete W_0\right).
\end{equation}
Similar to Sections~\ref{sec:eulermaruyama} and \ref{sec:milstein}, given the scheme $\phi$ as well as the approximate drift $\driftNN$ and diffusivity $\stdNN$ parametrized by $\weights$, we need to approximate the probability density $p_\weights(\X_{1}|\X_0,\deltat)$.
For schemes that rely on noise processes $\diffdiscrete W_0$ that can be sampled easily (e.g. Gaussian noise as in Euler-Maruyama and Milstein schemes), the key idea for our numerical approximation is the following. We begin by sampling a set of $M$ points $\diffdiscrete W_0^{(i)}$ from the distribution of $\diffdiscrete W_0$. Then, we use the SDE integration scheme $\phi$ to estimate candidates $\hat{\X}_{1}^{(i)}$ for the next point in time, given $\X_{0}$ and $\deltat$. Finally, we describe the density of these candidates through a mixture of Gaussians, each with a small standard deviation $\sigma_{\rm r}>0$, so that the actual probability distribution $p_\phi$ is approximated as
\begin{equation}\label{eq:numerical pdf approximation}
    p_\phi(\X_{1}|\X_{0},\deltat)\approx \frac{1}{M}\sum_{i=1}^M p_{\normal(\hat{\X}_{1}^{(i)},\sigma_{\rm r}^2)}(\X_1).
\end{equation}
The approximation accuracy of \eqref{eq:numerical pdf approximation} improves as $M\to\infty$ and $\sigma_{\rm r}\to 0$; for a discussion of this, we refer to standard literature on kernel density estimation~\cite{Higham-2001,schoelkopf-2018} and viscosity solutions for PDE~\cite{crandall-1983,crandall-1992}. \figref{fig:pdf approximation milstein} demonstrates this convergence for varying $\sigma$, with fixed $M=10^4$, $\X_0=0.01$, $\deltat=1.0$, $\drift(\X)=2\X$, $\std(\X)=\X$, and $\phi$ the Milstein scheme with the exact form for $p_\phi$ given through equation~\eqref{eq:milstein PDF}.
Many other density estimation methods are applicable here as well, such as generative-adversarial networks~\cite{goodfellow-2014} or normalizing flows~\cite{Kobyzev-2019}.
%
%

%
\review{
\subsection{Underdamped Langevin equations}\label{sec:langevin}
Up to this point, we discussed SDE where all coordinates are directly affected by noise. Second-order SDE of Langevin-type cannot be accurately described in this way, because the diffusivity is exactly zero for half of the coordinates:
\begin{subequations}\label{eq:langevin}\begin{align}
    \diff x_t&=v_t \diff t,\\
    \diff v_t&=\drift(x_t,v_t)\diff t+\std(x_t,v_t)\diff W_t.
    \end{align}
\end{subequations}
The density of $x_t$ is no longer Gaussian, and it would seem that we require a loss function other than~\eqref{eq:lossfunction EM} to approximate drift $\drift$ and diffusivity $\std$.
Equation~\eqref{eq:langevin} can be approximated using the (symplectic) Euler-Maruyama scheme
\begin{subequations}\label{eq:langevin EM}\begin{align}
    x_1&=x_0+v_0 \deltat,\\
    v_1&=v_0+\drift(x_1,v_0)\deltat+\std(x_1,v_0)\diffdiscrete W_0.
    \end{align}
\end{subequations}
This will not approximate the stationary distribution well, and other, more accurate schemes exist~\cite{mannella-2006}. Still, since we only assume access to snapshots $((x_0,v_0), (x_1,v_1), \deltat)$ and do not process longer time series, we can effectively view $x_1$ in~\eqref{eq:langevin EM} as a fixed parameter for the functions $\drift,\std$ over a time span of $\deltat$.
Then, $v_1$ is again normally distributed, conditioned on $x_1, v_0$ and step size $\deltat$, similar to~\eqref{eq:normal for xkp1}. Thus, loss function~\eqref{eq:lossfunction EM} can be used, {\textit{but now only for the dynamics of $v_t$}}, when given the corresponding values of $x_t$, to learn approximations $\driftNN$ and $\stdNN$. Including the symplectic adaptation which uses $x_1$ instead of $x_0$ to predict $v_1$, we thus must minimize
\begin{equation}\label{eq:lossfunction EM langevin}
    \lossfunction( \weights | \X_1,v_0,v_1,\deltat) := \frac{\left(v_{1}-v_0 - \deltat \driftNN(\X_1,v_0)\right)^2}{\deltat \,\stdNN(\X_1,v_0)^{2}} + \log \left|\deltat\stdNN \left(\X_1,v_0\right)^2 \right|+\log(2\pi).
\end{equation}
We show in \secref{sec:example Langevin} that using this adapted training scheme for data sampled from SDE of Langevin type~\eqref{eq:langevin} significantly improves approximation accuracy, compared to (incorrectly) using loss function~\eqref{eq:lossfunction EM} on $(x_t,v_t)$ data.
}

%
\review{
\subsection{Stochastic partial differential equations}\label{sec:spde}

Equation~\ref{eq:sde} defines a stochastic differential equation in finite-dimensional space, s.t. $x_t\in\mathbb{R}^n$. Stochastic partial differential equations (SPDEs) describe properties of functions embedded in infinite-dimensional spaces. In the context of agent-based systems, a description of the system behavior though an SPDE is useful if the number of agents is very large, but finite-size effects still play a role, so that a deterministic description though a deterministic PDE would be too restrictive. There are no generically applicable numerical schemes for SPDEs similar to the Euler-Maruyama scheme~\eqref{eq:em scheme}, and numerical methods of higher-order also typically do not exist~\cite{walsh-2006}. However, in certain cases, we can reformulate existing numerical schemes for specific SPDEs and still learn parts of its law in a black- or gray-box setting.
 For example, we can start with a stochastic wave equation that has both deterministic driving $\drift(x)$ and stochastic forcing $\std(x)$, and initial conditions $u_0,v_0$:
\begin{subequations}\label{eq:wave spde}
    \begin{align}
        \frac{\partial^2u}{\partial t^2} & = \frac{\partial^2u}{\partial x^2} + \drift(x) + \std(x)\diff W_t,\ x\in\mathbb{R}, t>0, \\
        u(x,0) & = u_0(x),\ \frac{\partial u}{\partial t}(x,0)=v_0(x);\ x\in\mathbb{R}.
    \end{align}
\end{subequations}
To solve \eqnref{eq:wave spde}, we choose a small step size $h>0$ and discretize space and time into points $x_i=ih$, $t_j=jh$. We choose a staggered grid on which we construct the solution, according to~\cite{walsh-2006}, as shown in \figref{fig:wave staggered grid}. The center grid points $(x_i,t_j)$ are excluded in the discretization, no solution is approximated on it.
\begin{figure}
    \centering
    \includegraphics[width=0.3\textwidth]{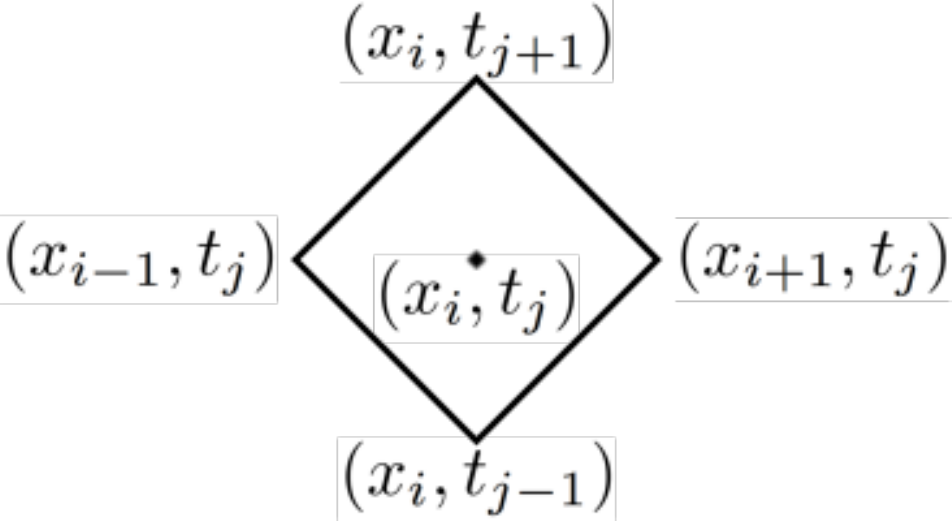}
    \caption{Staggered grid on which we can solve the stochastic wave equation. The center grid point $(x_i,t_j)$ is excluded in the discretization. The distance from $x_i$ to $x_{i+1}$ is the step size $h$, the same between $t_j$ and $t_{j+1}$.}
    \label{fig:wave staggered grid}
\end{figure}
The following numerical scheme then approximates the solution to the stochastic wave equation on the grid.
\begin{subequations}\label{eq:wave spde scheme}
    \begin{align}
        u_{i,-1}&=u(x_i,t_{-1})=\frac{1}{2}(u_0(x_{i-1})+u_0(x_{i+1}))-\deltat v_0(x_i),\ i\text{ odd};\\
        u_{i,1}&=u_0(x_{i-1})+u_0(x_{i+1})-u_{i,-1}\nonumber\\
        &+\frac{\deltat^2}{2}\drift(x_i)
        +\frac{1}{2}\std(x_i)\diffdiscrete W_0(\deltat^2/2), i\text{ odd};\\
        u_{i,j+1}&=u_{i+1,j}+u_{i-1,j}-u_{i,j-1}\nonumber\\
        &+\deltat^2 \drift(x_i)
        +\frac{1}{2} \std(x_i)\diffdiscrete W_0(\deltat^2),\ (i(j+1))\text{ even}, j\geq 2.
    \end{align}
\end{subequations}
The scheme~\eqref{eq:wave spde scheme} can be reformulated to match the Euler-Maruyama scheme~\eqref{eq:em scheme} by defining 
\begin{equation*}
        \hat{\deltat}:=\frac{1}{2}\deltat^2,\ 
        \hat{u}_{i,j}:=\frac{1}{2}(u_{i+1,j}+u_{i-1,j}),\  \tilde{u}_{i,j+1}:=\frac{1}{2}(u_{i,j+1}+u_{i,j-1}).
\end{equation*}
Then,
\begin{equation*}
    \tilde{u}_{i,j+1}=\hat{u}_{i,j}
        +\hat{\deltat} \drift(x_i)
        +\std(x_i)\diffdiscrete W_0(\hat{\deltat}),\ (i(j+1))\text{ even}, j\geq 2,
\end{equation*}
which is the same scheme as \eqref{eq:em scheme}. Given data $\trainingdata=\lbrace \tilde{u}_{i,j+1}, \hat{u}_{i,j}, \hat{\deltat}\rbrace_{i,j}$, we can learn the forcing terms $\drift, \std$ through minimization of the loss~\eqref{eq:lossfunction EM}.
In this manuscript, we only considered the autonomous case, where $\drift$ and $\std$ do not depend on time.
Note that the SPDE~\eqref{eq:wave spde} is also of Langevin type, and the reformulation of the Euler-Maruyama scheme here corresponds to the reformulation explained in \secref{sec:langevin}.
}

%
\subsection{Exploiting latent spaces}\label{sec:latent spaces}
%
While we initially consider cases where the coarse variables are already known (e.g. their mean field or quasi-chemical approximation variables), many observations in the real world do not immediately reveal the smallest dimensionality of the space in which the dynamics evolve.
\review{Latent space identification and related unsupervised learning are possible with a wide variety of techniques, also for SDE~\cite{hasan-2022}. In our work,} we learn such latent spaces with Diffusion Maps~\cite{coifman-2006} as well as an SDE-integrator-informed auto-encoder with the SDE loss~\eqref{eq:lossfunction EM} augmented by the log mean squared loss:
$\lossfunction(\weights|\trainingdata)=\log(\frac{1}{\numpoints}\sum_{\idx=1}^\numpoints(\text{decoder}(\text{encoder}(x_0^{(\idx)}))-x_0^{(\idx)})^2)+\lossfunction_{\text{Euler-Maruyama}}$.
%
The Diffusion Map coordinates define a latent space; Learning an SDE on these new variables, and then mapping back to the original data is thus similar to latent SDE identification~\cite{li2020scalable}. 
With Diffusion Maps, we first identify coarse variables and then learn the effective SDE. This has benefits compared to an end-to-end auto-encoder approach, because by construction, the latent coordinates are invariant to isometry and sampling density in the ambient space~\cite{coifman-2006}. Learning the latent coordinates together with the drift and diffusivity functions instead leads to a latent space that is adapted to the specific SDE. This construction usually does not have the same invariance properties as the coordinates from Diffusion Maps.
A description of the methods to obtain these latent spaces and the identification of SDE on them for kMC lattices can be found in the supplemental material, \ref{sec:autoencoder}.

\section{Computational experiments}\label{sec:illustrative examples}
We first illustrate the neural network identification techniques on toy examples, and then show an application to learning effective SDEs from lattice dynamics simulations. Tab.~\ref{tab:experiment parameters} lists important training parameters.
All computational experiments were performed on a single desktop computer with an AMD Ryzen 7 3700X 8-Core Processor, 32GB RAM, and a NVIDIA GeForce GTX 1070 GPU. Training did not take longer than 30 minutes for each case, data generation did not take longer than two hours. The neural networks and loss functions were defined in TensorFlow V2.4.1~\cite{tensorflow2015-whitepaper} (Apache 2.0 license).
For the toy examples, we created snapshots 
by integrating the SDEs with 10 steps of \eqnref{eq:em scheme} from $t=0$ to $t=\deltat$, only storing the initial and final result.
The given network topology was used twice, separately for drift and diffusivity. If not otherwise stated, we use a test/validation split of 90\%/10\%, the ADAM optimizer with standard parameters \texttt{learning\_rate}=$0.001$, ${\beta_1}=0.9$, ${\beta_2}=0.999$, ${\epsilon}=10^{-7}$, batch size of 32, 100 epochs, and exponential linear unit activation functions.

\begin{table}[ht]
    \centering
    \caption{Parameters, equations for loss, training and validation errors for the experiments.}
    \begin{tabular}{lcccccc}
    \toprule 
       \textbf{Example}  & $\deltat$ & \textbf{\# points} & \textbf{Network topology} & \textbf{Loss} & \textbf{Train} & \textbf{Val.} \\
       \midrule 
        \secref{sec:example cubic}: cubic, 1D &  0.01 & 10,000 & $1|50|50|1$ & \eqref{eq:lossfunction EM} & -0.00354 & -0.0111 \\
        \secref{sec:example cubic}: cubic, 1D &  0.01 & 10,000 & $1|50|50|1$ & \eqref{eq:milstein PDF} & -0.0010 & -0.0112 \\
        \secref{sec:example cubic}: linear, 3D &  $0.25$ & 15,000 & $3|50|50|3/6$ (drift) & \eqref{eq:lossfunction EM} & -7.60 & -7.63\\
         &   &  & $3|50|50|6$ (diff.) &  &  & \\
        \secref{sec:example cubic}: 1D--19D &  $0.01$ & 10,000 & $\ndim|50|50|\ndim$ & \eqref{eq:lossfunction EM} & varies & varies\\
        \secref{sec:sde from particles}a: Gillespie & varies & 90,000 & $2|50|50|2$ & \eqref{eq:lossfunction EM} & -10.9 & -11.0 \\
        \secref{sec:sde from particles}b: kMC lattice  & varies & 90,000 & $2|50|50|2$ & \eqref{eq:lossfunction EM} & -11.7 & -11.8\\
        \review{\secref{sec:example Langevin}: Langevin} & $0.05$ & 10,000 & $2|50|50|2$ & \eqref{eq:lossfunction EM langevin} & -0.44 & -0.42\\
        \review{\secref{sec:example wave SPDE}: wave SPDE} & $0.001$ & varies & $2|50|50|2$ & \eqref{eq:lossfunction EM} & varies & varies
        \\ \bottomrule
    \end{tabular}
    \label{tab:experiment parameters}
\end{table}

\subsection{Toy examples: cubic drift, linear diffusivity, higher dimensions}\label{sec:example cubic}
\begin{figure}[ht]
    \centering
    \includegraphics[height=0.15\textheight]{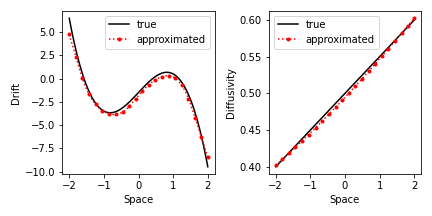}
    \includegraphics[height=.15\textheight]{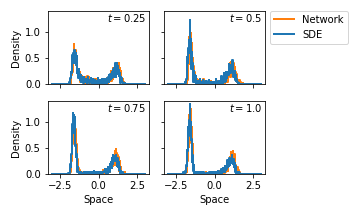}
    \caption{\label{fig:example_1dcubic_functions} Approximation vs. true values of $\drift$ and $\std$ in example~\ref{sec:example cubic}. Densities over time, from points sampled uniformly in $[-2, 2]$ and then integrated using the true SDE and using the approximation of drift and diffusivity with the network.}
\end{figure}

We tested our implementation with examples in one dimension first, with drift $\drift$ and diffusivity $\std$ defined through
$\drift(\X_t)=-2\X_t^3 - 4 \X_t+1.5,\ \std(\X_t)=0.05 \X_t + 0.5$.
A comparison between the learned and true functions is shown in \figref{fig:example_1dcubic_functions}, along with densities obtained from sampling the true and approximate SDE. We trained networks with both Euler-Maruyama and Milstein loss functions, with very comparable results---even the training and validation curves were very similar. When increasing the dimension from $\ndim=1$ to $\ndim=19$, approximation quality decreases after $\ndim=6$.
Non-diagonal diffusivity matrices were approximated accurately for $\ndim=3$ (see supplemental material).
%

\subsection{Learning coarse, effective SDE from fine-grained (particle) simulations}\label{sec:sde from particles}

We validate our approach through a well-studied model for which analytical approximations at several levels of coarse-graining are available. The problem is the dynamics of epidemics on a lattice, which we briefly outline below. We consider two different versions of the model, and ``learn'' the corresponding effective SDEs. The first version at the finest scale involves Monte Carlo lattice simulations; the second version involves Gillespie simulation of the mechanistic scheme implied by the mean field model.
Not surprisingly, if we use mean field SDE data, we discover an approximation to the mean field SDE. However, we find an approximation to dynamics in the mean field variables which is different than the mean field SDE, if we get data from Gillespie/lattice simulations and coarse-grain it in terms of mean field variables~\cite{makeev-2002,makeev-2002b}.

\textbf{Epidemic SIRS model:}
We consider a hierarchical sequence of three stochastic models describing the dynamics of the classical SIR-type (susceptible-infected-recovered) epidemic process~\cite{kermack-1927,Siettos-2013,allen-2017}. The first, most detailed model is represented by kinetic Monte Carlo (kMC) simulations on a lattice. The kMC model considers a two-dimensional square lattice of size $N=N_x\times N_y$ with periodic boundary conditions, where each site contains a single individual. In accordance to the health state, each individual can be in one of three states: Susceptible (S), Infected (I) or Recovered (R). The elementary reactions (independent Poisson processes) occurring on a lattice are as follows: 1) $I + S \to I + I$ (rate constant $k_1$; infection), 2) $I \to R$ (rate constant $k_2$; recovery with immunity), 3) $R \to S$ (rate constant $k_3$; loss of immunity).

The infection spreads through nearest neighbor bonds only; thus the first reaction may occur only when I and S are the first neighbors on a lattice. Elementary events take place with transition probabilities $k_p$ ($p=1,2,3$). When $k_3>0$, the model is called as SIRS, while the SIR model is a particular case of the SIRS model with $k_3=0$. In addition to steps 1)-3), migration of individuals occurring via the exchange mechanism is considered: 4) $S_i + I_j \to I_i + S_j$ (rate constant $\migrationrate_1$), and 5) $S_i + R_j \to R_i + S_j$ (rate constant $\migrationrate_2$). Here, the sites $i$ and $j$ are first neighbors on a lattice.
The time evolution of the probability distribution for the populations of individuals on a lattice can be described by the chemical master equation, derived from the Markov property of the underlying stochastic process. To generate realizations of this process, we applied the so-called rejection-free direct kMC algorithm~\cite{Bortz-1975,Gillespie-1976} which produces a sequence of configurations and the times when they change from one to the other.
It is interesting to mention that the same lattice kMC model can be formulated in a mathematically equivalent form using the following three types of states: Excited (infected), Refractory (recovered) or Quiescent (susceptible).
In this case the model may reproduce the typical patterns of excitable media such as traveling pulses and spiral waves~\cite{Makeev-2017}.
Furthermore, we assume that $\migrationrate=\migrationrate_1=\migrationrate_2$. In essence, the migration rate $\migrationrate$ may take into account quarantine restrictions and other regulations in place to keep the infection from spreading. Small $\migrationrate$ values correspond to strong restrictions. If a spatially homogeneous, random distribution of all individuals on a lattice is assumed, the problem can be significantly simplified. Such a homogeneous distribution can be achieved in the ``fast migration'' limit $(\migrationrate\to\infty)$.

The second stochastic model we study is represented by the Gillespie's stochastic simulation algorithm (SSA,~\cite{Gillespie-1976}) that is used to simulate a well-stirred, spatially homogeneous system. For the SIRS model, SSA operates in terms of three integer stochastic variables: the number of susceptible individuals, $n_0$, the number of infected  individuals, $n_1$, and the number of recovered individuals, $n_2$; their concentrations are calculated as: $\SIRvar_p=n_p/\numparticles$. The mass balance implies that $\SIRvar_0+\SIRvar_1+\SIRvar_2=1$. At each time step of the SSA algorithm the following three rates are calculated through $r_1=4k_1 \SIRvar_0 \SIRvar_1,\ r_2=k_2 \SIRvar_1,\ r_3=k_3 \SIRvar_2.$
Then, an event to be performed is selected with a probability proportional to its rate and the numbers $n_p$ are updated accordingly.

The third stochastic model is represented by a set of stochastic differential equations (SDE) of the Langevin type which, under certain assumptions, approximates the results of the SSA algorithm~\cite{Gillespie-2000}. The SDE model at finite $N$ is as follows:
\begin{subequations}\label{eq:SIR SDE model}\begin{align}
\frac{\diff\SIRvar_0}{\diff t}&=-r_1+r_3-\sqrt{r_1/N}  \diff W_1 (t)  + \sqrt{r_3/N}  \diff W_3 (t), \\
\frac{\diff\SIRvar_1}{\diff t}&=(r_1-r_2 )  + \sqrt{r_1/N}  \diff W_1 (t)  - \sqrt{r_2/N}  \diff W_2 (t),\\
\frac{\diff\SIRvar_2}{\diff t}&=r_2-r_3+\sqrt{r_2/N}  \diff W_2 (t)- \sqrt{r_3/N}  \diff W_3 (t). 
\end{align}
\end{subequations}
Here, $W_p(t)$ are independent Wiener processes such that $\diff W_p (t)\sim\normal(0,\diff t)$. Only two of three SDEs must be considered, since $\SIRvar_0+\SIRvar_1+\SIRvar_2=1$. In the limit $N\to\infty$, the SDE system transforms to the standard mean field SIRS model. 
In particular, for the SIR model this SDE system reduces to having a diagonal diffusivity matrix: 
\begin{equation}\label{eq:SIR mean field}
\diff\SIRvar_0=-r_1 \diff t+\sqrt{r_1/N}  \diff W_1 (t),                               \ 
\diff\SIRvar_2 =r_2 \diff t+\sqrt{r_2/N}  \diff W_2 (t).
\end{equation}
\eqnref{eq:SIR SDE model} or \eqref{eq:SIR mean field} can be integrated using the Euler-Maruyama method with fixed time step $\deltat$. The SDE models approximate the solution of both SSA and kMC models.
The following parameters and initial conditions are used in the simulations:  $k_1=1$, $k_2=1$, $k_3=0$; $\SIRvar_0(0) = 0.9$, $\SIRvar_1(0) = 0.1$.
In the limit of infinite $N$, the mean field equations simplify to
\begin{equation}\label{eq:SIR mean field continuous}
\frac{\diff\SIRvar_0}{\diff t}=-k_1\SIRvar_0\SIRvar_1,\ 
\frac{\diff\SIRvar_1}{\diff t}=k_1\SIRvar_0\SIRvar_1-k_2\SIRvar_1.
\end{equation}

We assume that SSA and kMC simulations generate trajectories which can be approximated by the solution of two latent SDEs with diagonal diffusion matrix, similar to \eqnref{eq:SIR mean field}. Under this assumption, the neural network was used to learn the unknown drift and diffusion terms at the fixed values of the model parameters $k_1, k_2, k_3$ and $\numpoints$. Training was performed using a set of $\numpoints_{tr}$ short-time stochastic simulations on a $N_1\times N_2$ lattice with the random, but physically-relevant initial conditions for the average concentrations $y_m\in(0,1)$, $m\in\left\lbrace 0,1,2\right\rbrace$.
The initial species distribution on a lattice was also random. During each independent SSA and kMC simulation run, the data points were evaluated at the time moments $t_0\in\left\lbrace 0, t_1, \dots, t_{\text{max}}\right\rbrace$, where $\deltat_i=t_{i+1}- t_i \approx \deltat$.
The values of $t_{\text{max}}$ and $\deltat$ are the fixed parameters of the algorithm.
Note that a time step in the SSA/kMC algorithm is a random variable. The time is advanced via an increment given from the exponential distribution: $dt=-\log(\xi)/R$,
where $R$ is the total reaction rate and $\xi\in(0,1)$  is a uniformly distributed random number. Therefore, $\deltat_i$ are close to $\deltat$, but they are not constant.
Each data point used for training contains the following five values:
$\deltat_i, y_{(0,i)}, y_{(2,i)}, y_{(0,i+1)}, y_{(2,i+1)}$. The total number of data points is $\numpoints_{dp}=\numpoints_{tr} \left([t_{\text{max}}/\deltat]-1\right)$. The trained network approximates the unknown drift and diffusion terms at any given values of $y_m\in(0,1)$. Thus, the SDE approximation in combination with the Euler-Maruyama method provides a time-stepper (decoder) which can reproduce the dynamics. 



\begin{figure}[ht]
    \centering
    \includegraphics[width=0.3\textwidth]{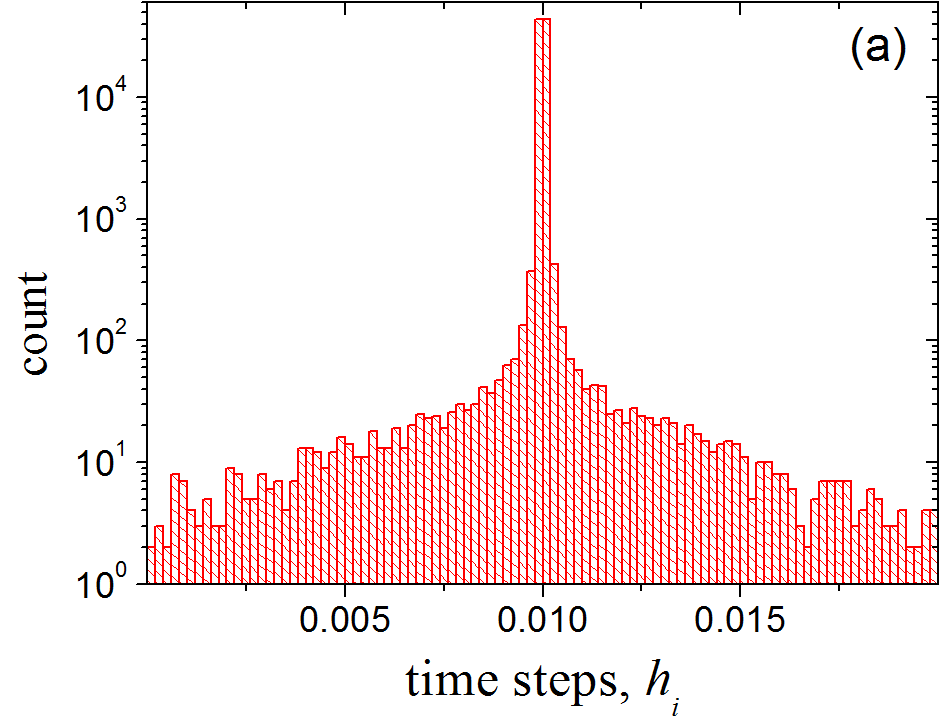}
    \includegraphics[width=0.3\textwidth]{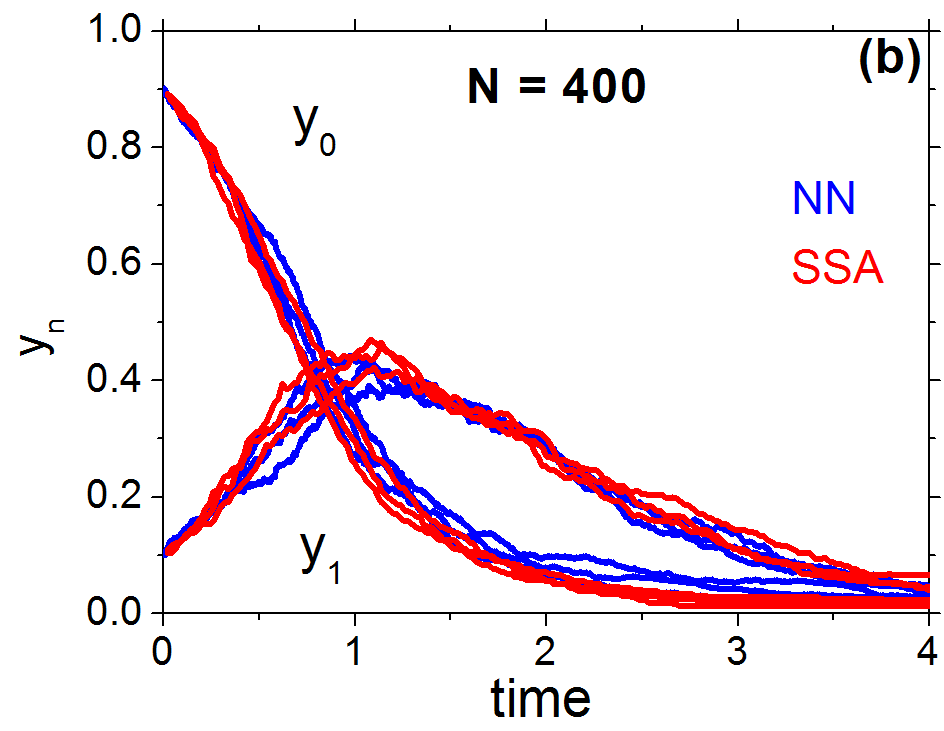}
    \includegraphics[width=0.3\textwidth]{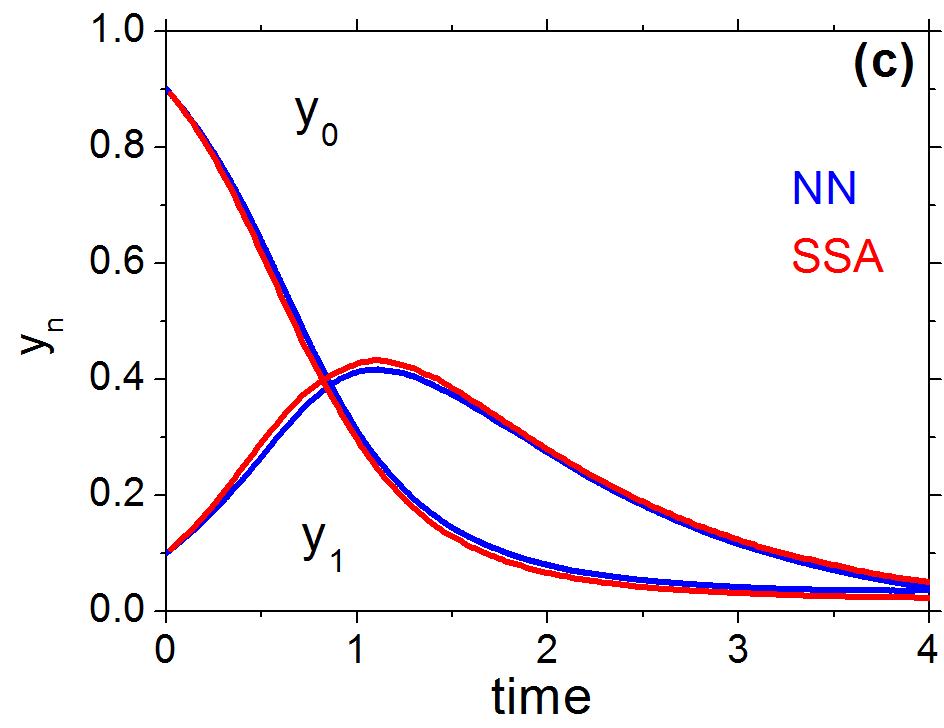}
    \caption{Histogram of the time steps $\deltat_i$ used in the training data set (a). Three sample paths of $\SIRvar_0$ and $\SIRvar_1$ (b); Average of $200$ paths  with deviation of $\max |\SIRvar_{i,\text{NN}} - \SIRvar_{i,\text{SSA}}|<0.02$ (c). Blue (red) lines correspond to NN (SSA) results. Parameters: $\numpoints=400$; $\numpoints_{tr}=10000$, $t_{\text{max}}=0.1$, $\deltat=0.01$.}
    \label{fig:gillespie results}
\end{figure}

\paragraph{Gillespie simulations:}\label{sec:Gillespie}
\figref{fig:gillespie results} shows that a network can be trained to match the results of SSA simulations. A distribution function (histogram) of the time steps $\deltat_i$ is plotted in \figref{fig:gillespie results}(a). It demonstrates that our approach works with variable time steps in the training data set. Three sample paths of $\SIRvar_0$ and $\SIRvar_1$ and the average of $200$ paths are shown in Figs.~\ref{fig:gillespie results}(b) and~\ref{fig:gillespie results}(c).

\begin{figure}[ht]
    \centering
    \includegraphics[width=1\textwidth]{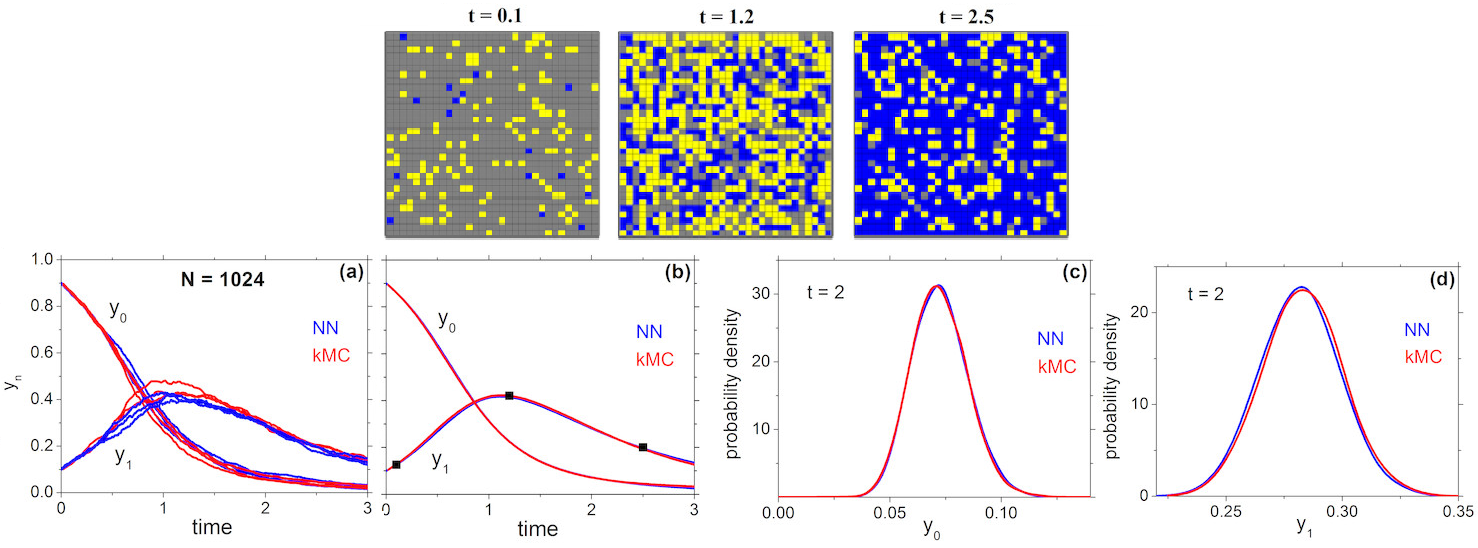}
    \caption{Illustration of the kMC lattice (top row), physically relevant values $\SIRvar_0, \SIRvar_1$ measured over time resp. simulated with the identified SDE from the network (a), averaged paths over 200 simulations with a deviation $\max |\SIRvar_{i,\text{NN}} - \SIRvar_{i,\text{kMC}}|<0.01$ (b), and propagated densities from the initial condition until $t=2$ (c,d). Lattice snapshots show S, I and R-type species as grey, yellow and blue squares, respectively. Parameters: $32\times32$ lattice, $\migrationrate=50$; $N_{tr} = 4000$, $t_{\text{max}} = 0.05$, $\deltat = 0.01$.}
    \label{fig:kmc_lattice_illustration}
\end{figure}

\paragraph{Kinetic Monte-Carlo lattice simulations:}\label{sec:kMC lattice}
\figref{fig:kmc_lattice_illustration} shows that our approach successfully reproduces the results of kMC simulations on a $32\times 32$ lattice.
%
%
The Euler-Maruyama method with a fixed time step $\deltat_{\text{EM}}=0.002$ was used in numerical integration of the network SDE approximation. \figref{fig:kmc_lattice_illustration}(a) demonstrates three sample paths obtained using the neural network (blue) and kMC (red, ``true solution'').
The averaged paths are presented in \figref{fig:kmc_lattice_illustration}(b). A sufficiently high accuracy network approximation is demonstrated in Figs.~\ref{fig:kmc_lattice_illustration}(c) and~\ref{fig:kmc_lattice_illustration}(d), where the probability density functions  for $\SIRvar_0$ and $\SIRvar_1$ at $t=2$ are shown. They were estimated by NN and kMC, and are visually identical.

The described examples consider the lattice SIR model which has only the trivial steady states with $y_1=0$. When $k_3>0$, the system exhibits richer dynamic behavior. Simulations with $k_3=0.2$ were performed as well, and the network model is capable of reproducing the correct steady state and damped oscillations (the mean field SIRS model has a focus-type steady state at these values of parameters).
%
%
This is actually surprising: When the SIRS model is considered, the diffusion matrix is not diagonal, because there are three different reactions. Nevertheless, the trained network is capable here of approximating the kMC results, even when only trained with a diagonal diffusivity matrix.

\review{
\subsection{Learning underdamped Langevin dynamics}\label{sec:example Langevin}
The reformulation of the loss function~\eqref{eq:lossfunction EM} to \eqref{eq:lossfunction EM langevin} introduced in \secref{sec:langevin} allows us to also learn non-Gaussian dynamics of Langevin-type. We will now demonstrate this on a simple example, and show that incorporating the assumption of Langevin-type dynamics for data from Langevin systems is necessary for accurate approximation of drift and diffusivity functions. The system we investigate is defined through one-dimensional state $x_t$ and velocity $v_t$, such that
\begin{equation}\label{eq:langevin example}\left.\begin{array}{rcl}
    \diff x_t&=&v_t \diff t,\\
    \diff v_t&=&\drift(x_t)\diff t+\std \diff W_t,
    \end{array}\right\rbrace
\end{equation}
where $\gamma=\frac{1}{2}$, $\beta=10$, $\drift(x_t)=-\nabla U(x_t)-\gamma v_t$ with potential $U(x_t)=\frac{1}{4}x_t^4+\frac{1}{2}x_t^2$, and $\std =\sqrt{2 \gamma / \beta}$.
When approximating $\drift$ and $\std$, we assume they depend on both $x_t$ and $v_t$, which creates a more challenging learning problem. As a loss function, we use \eqref{eq:lossfunction EM langevin}. The data from the spatial coordinate $x_t$ is used as a fixed parameter for each given snapshot $((x_0,v_0), (x_1,v_1), \deltat)$. The result is shown in \figref{fig:example 8 paths}, with probability densities for different times shown in \figref{fig:example 8 densities}.
\begin{figure}[ht]
    \centering
    \includegraphics[height=0.15\textheight]{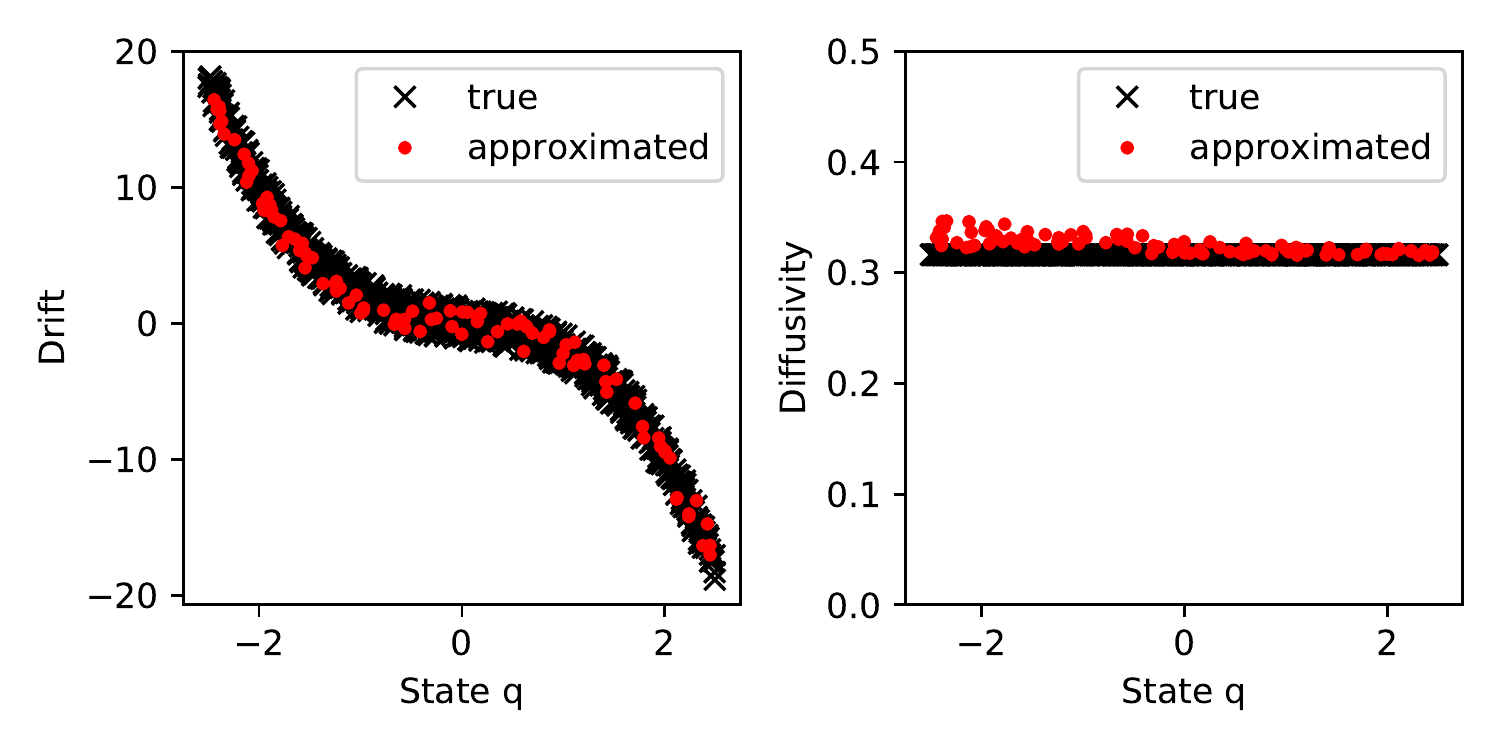}
    \includegraphics[height=0.15\textheight]{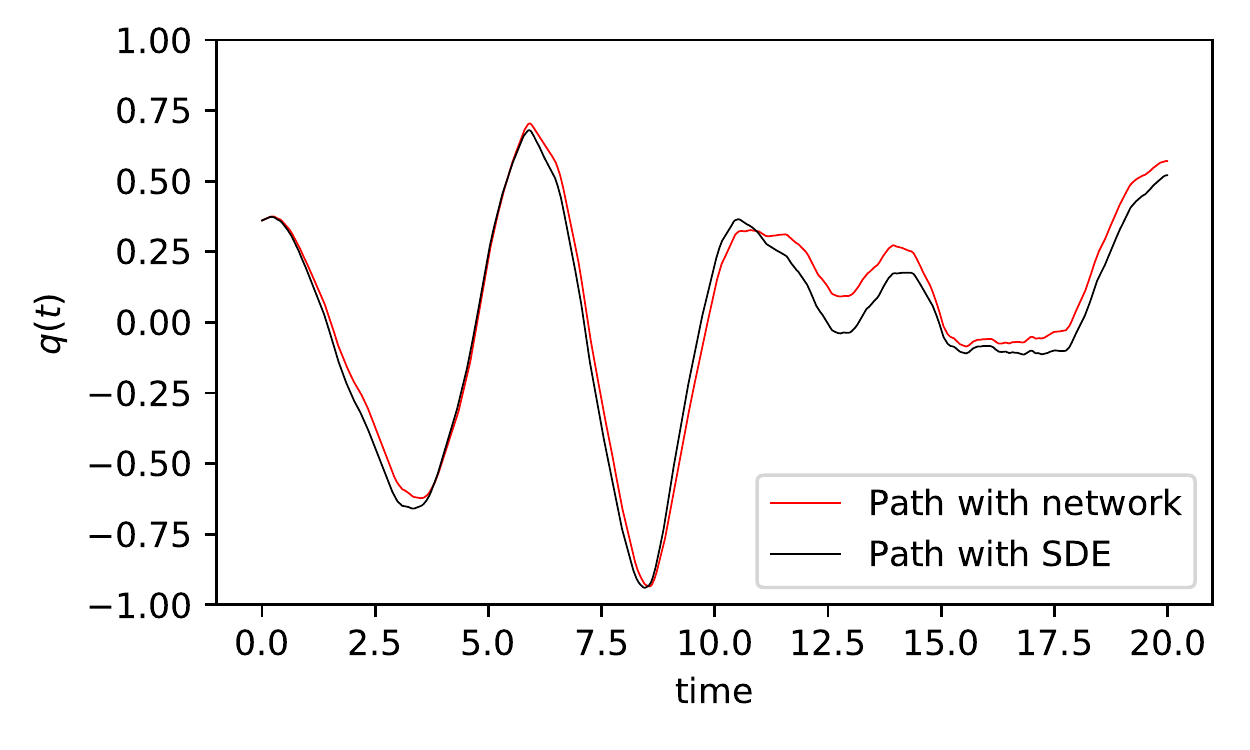}
    \caption{\review{Approximations of drift and diffusivity, and sample path comparison when using the same sample path of the noise process $W_t$. The spread of values in the drift term is due to the small influence of $-\gamma v_t$ in the true function $\drift$, which is also captured accurately.}}
    \label{fig:example 8 paths}
\end{figure}
\begin{figure}[ht]
    \centering
    \includegraphics[height=0.15\textheight]{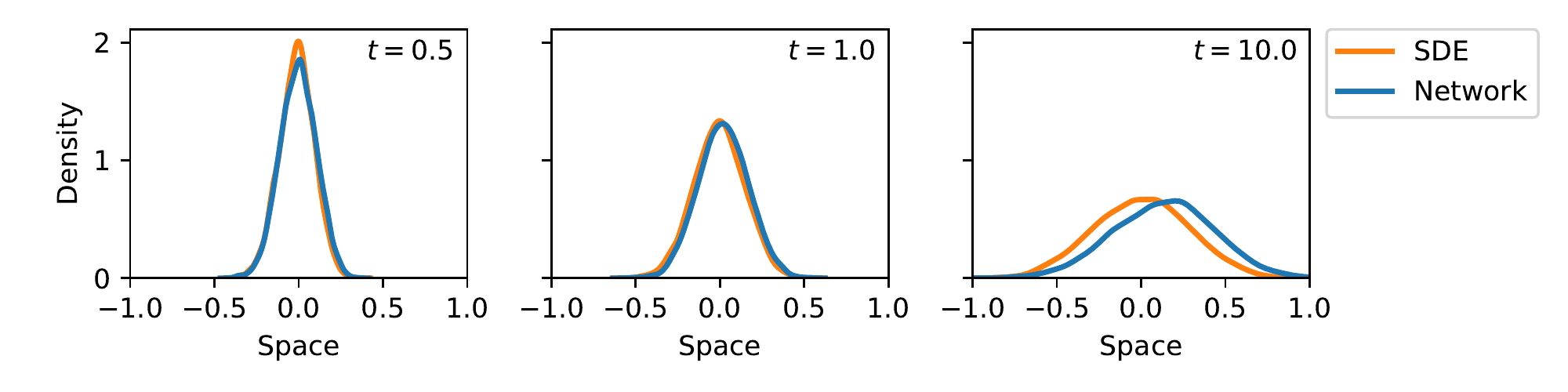}
    \caption{\review{Probability density functions at times $t=0.5, 1.0, 5.0$, starting with $x~\sim\normal(0,\frac{1}{10})$, $v=0$. The network approximation works well for smaller times, but the integrated error accumulates for larger ones.}}
    \label{fig:example 8 densities}
\end{figure}

When using loss function \eqref{eq:lossfunction EM} for both dimensions $(x_t,v_t)$ instead, we effectively (incorrectly) assume the SDE \eqref{eq:sde} instead of \eqref{eq:langevin}. In this case, the training fails to converge, as shown in \figref{fig:example 8 agnostic paths}, even when trained with 20 times the number of epochs. The final loss value in this experiment is $-8.4$, and would diverge to $-\infty$ with increasing number of data points. The issue is also apparent in the loss evaluated during training, as shown in~\figref{fig:example 8 loss}.
\begin{figure}[ht]
    \centering
    \includegraphics[height=0.15\textheight]{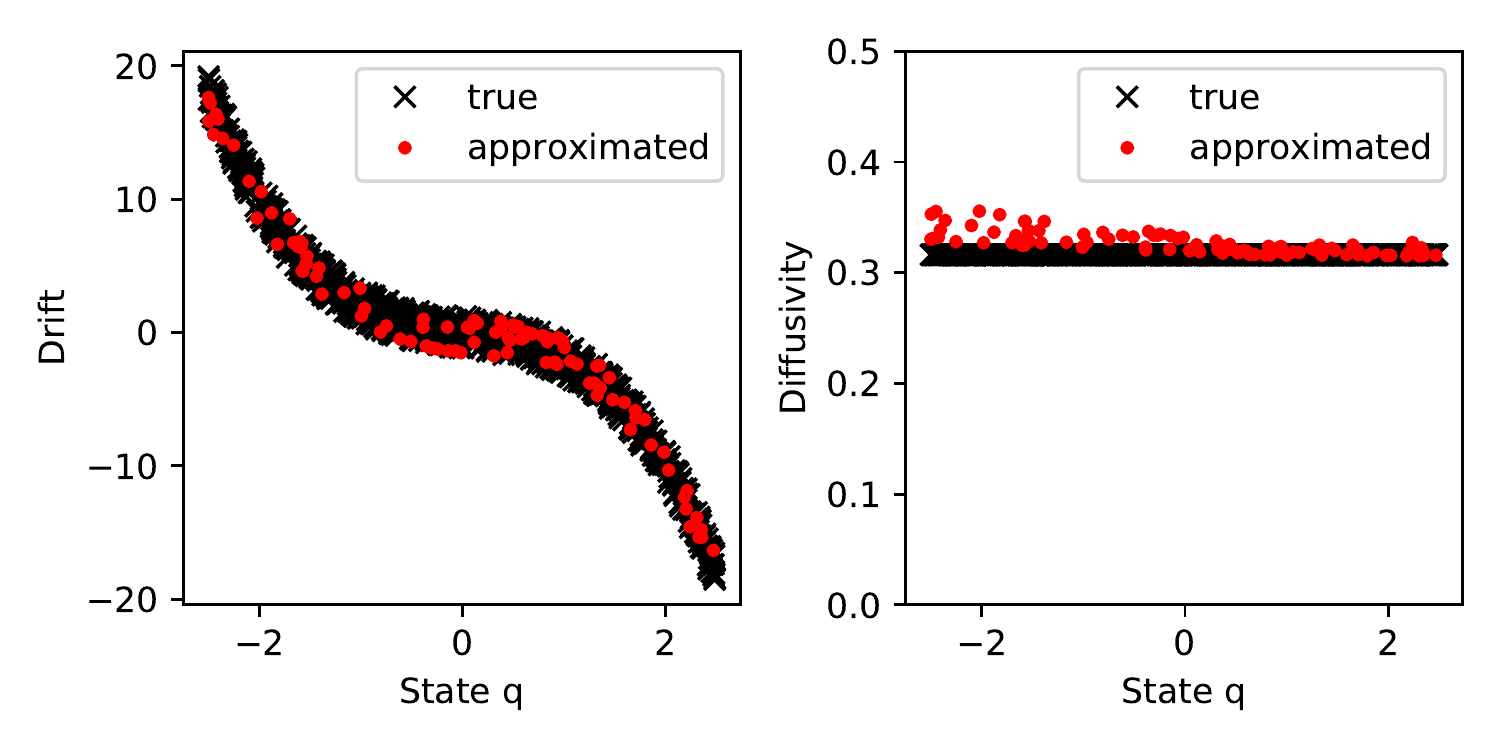}
    \includegraphics[height=0.15\textheight]{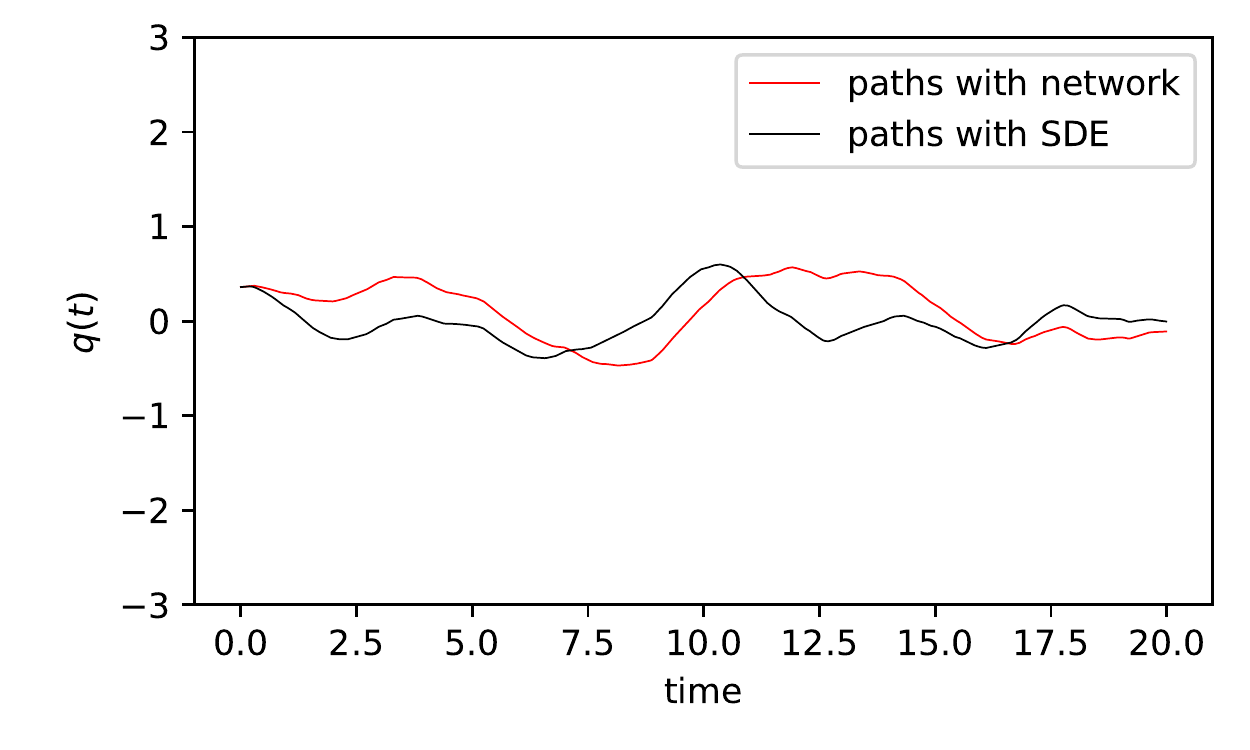}
    \caption{\review{Approximations of drift and diffusivity, and sample path comparison when using the same noise process, if the wrong type of SDE is assumed. Drift and diffusivity terms for $v_t$ are only learned after 20 times more training time, and the path prediction also fails.}}
    \label{fig:example 8 agnostic paths}
\end{figure}
\begin{figure}[ht]
    \centering
    \includegraphics[height=0.15\textheight]{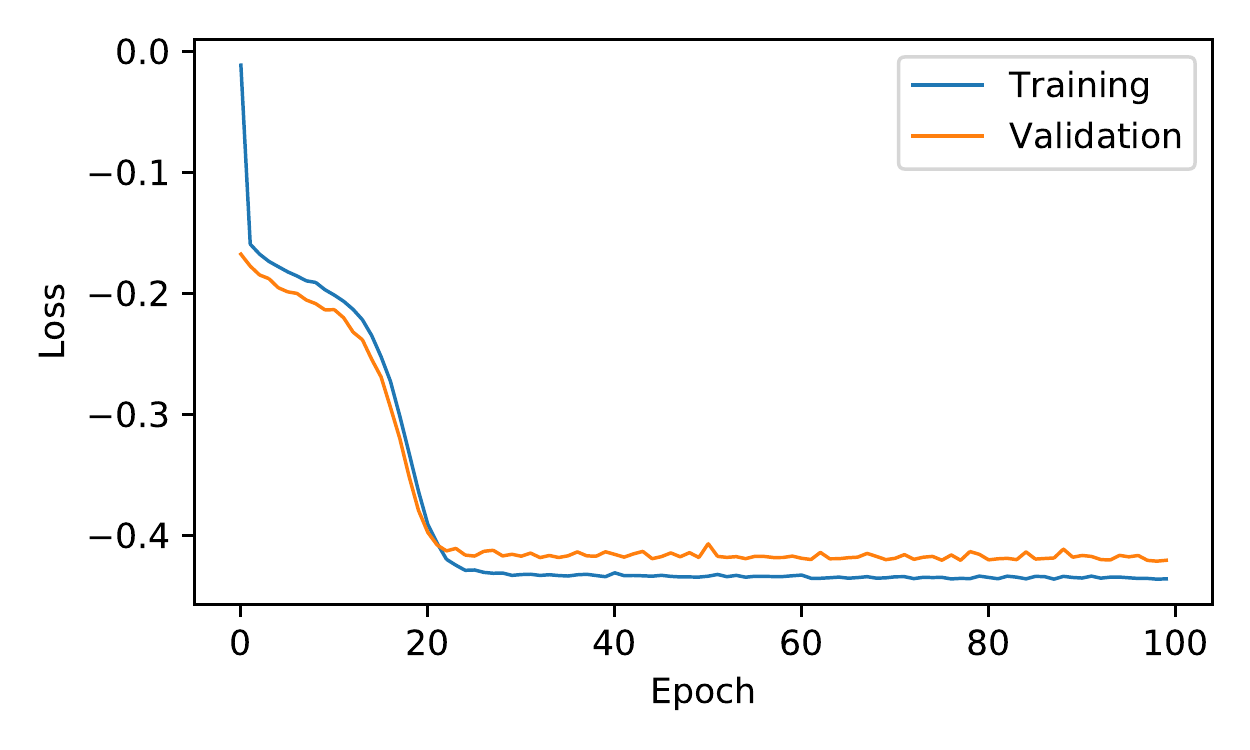}
    \includegraphics[height=0.15\textheight]{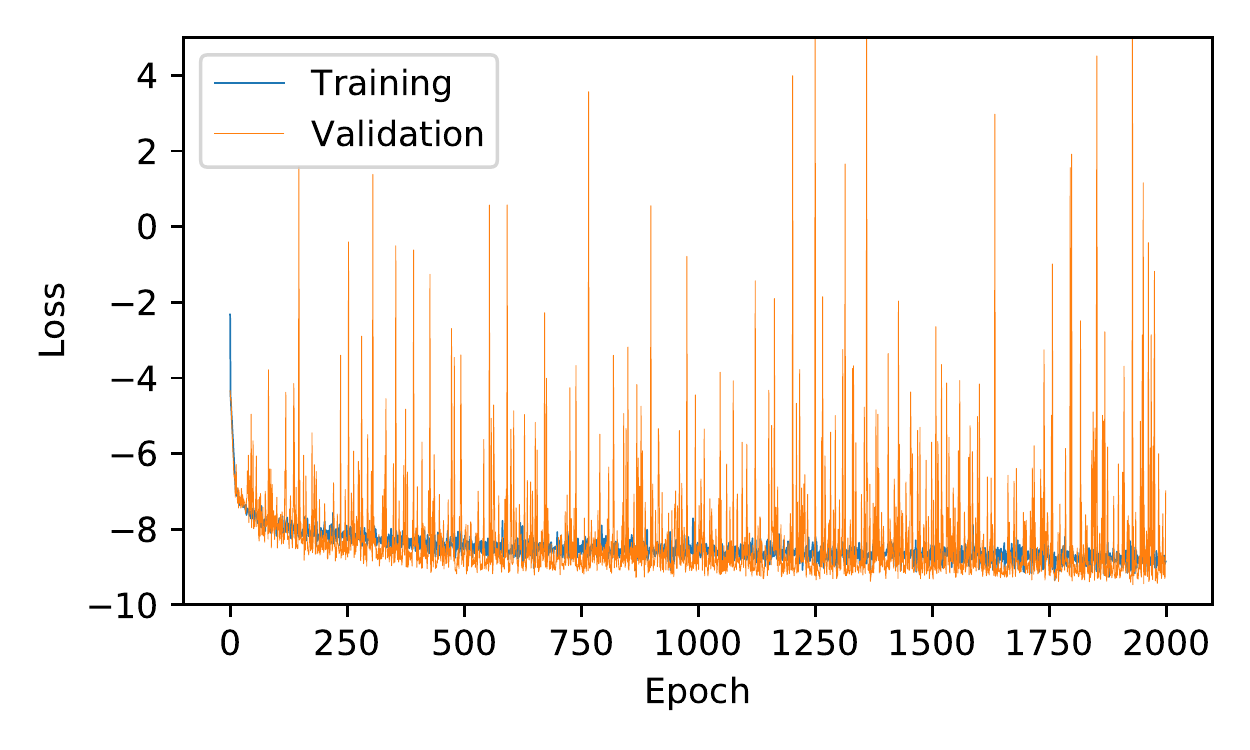}
    \caption{\review{Loss against epoch when training with correct Langevin loss \eqref{eq:langevin} and incorrect Euler loss~\eqref{eq:lossfunction EM}. Note the erratic behavior of the loss, and that training 20 times longer also does not lead to proper convergence. The numerical value of the loss in the right panel is also much lower, resulting from an approximated diffusivity very close to zero.}}
    \label{fig:example 8 loss}
\end{figure}
}

\review{
\subsection{Learning forcing terms in SPDEs}\label{sec:example wave SPDE}

We now demonstrate learning deterministic driving, $\drift(x)$, and stochastic forcing, $\std(x)$, terms in a stochastic PDE, using the reformulation of the numerical scheme \eqref{eq:wave spde scheme} with step size $\deltat=0.001$.
The training data in \figref{fig:example 7 wave pde} were generated using autonomous, spatially varying forcing terms,
\begin{equation*}
    \drift(x)=5\sin(4 \pi x),\ \std(x)=1/20 (1 + \exp(-150 (x-1/2)^2)).
\end{equation*}
We use our methodology to identify approximations of these two functions with $N=500,000$ pairs of training data. In addition to this experiment, \figref{fig:example 7 wave pde loss experiment} shows how the network approximations converge to the true functions (which would be unknown in a real setting), when the number of available data points (triples $(u_n,u_{n+1},\deltat)$) is increased while keeping the number of network parameter updates in one training run constant at approximately $10^5$. We perform ten training runs per data set size to illustrate variability of the results due to stochastic training and data point selections. The loss~\eqref{eq:lossfunction EM} used during training and validation is approximately constant for all data sets of size $>10^4$, indicating convergence of training for all cases where the number of data points surpasses the number of trainable parameters (including weights and biases, we use a network with 5602 trainable parameters).
\begin{figure}[ht]
    \centering
    \includegraphics[height=0.15\textheight]{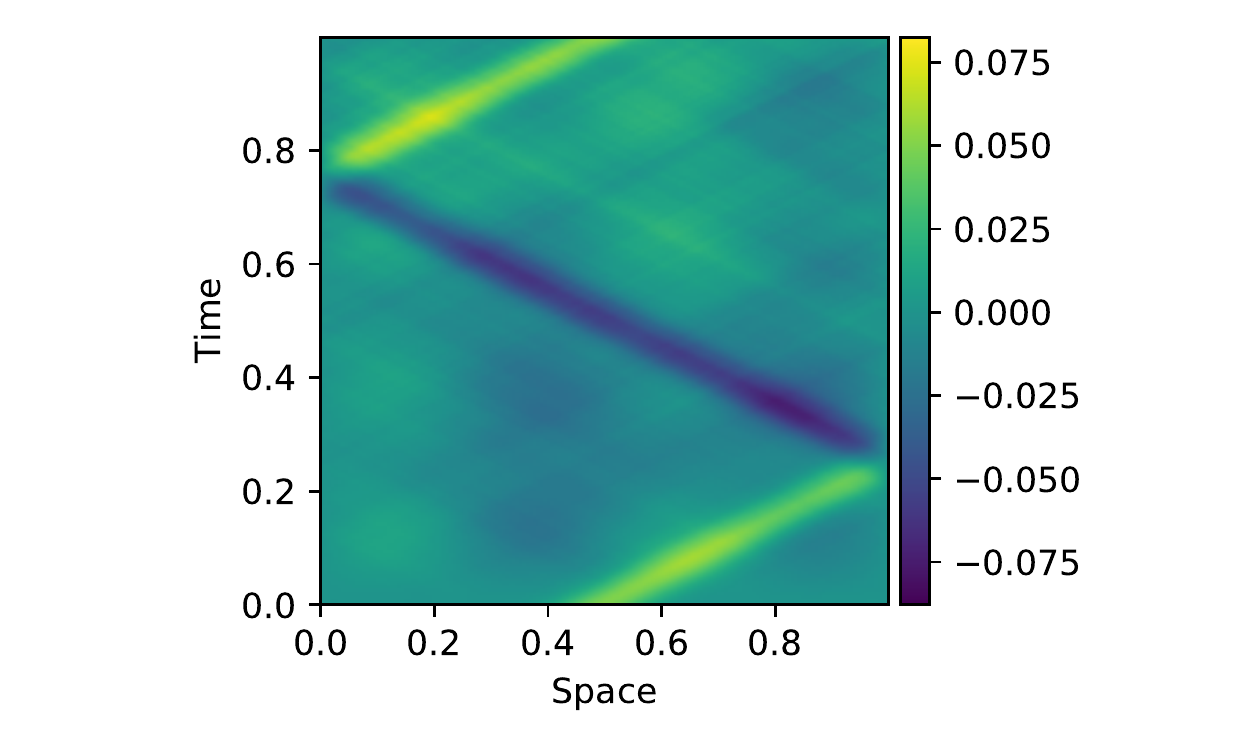}
    \includegraphics[height=0.15\textheight]{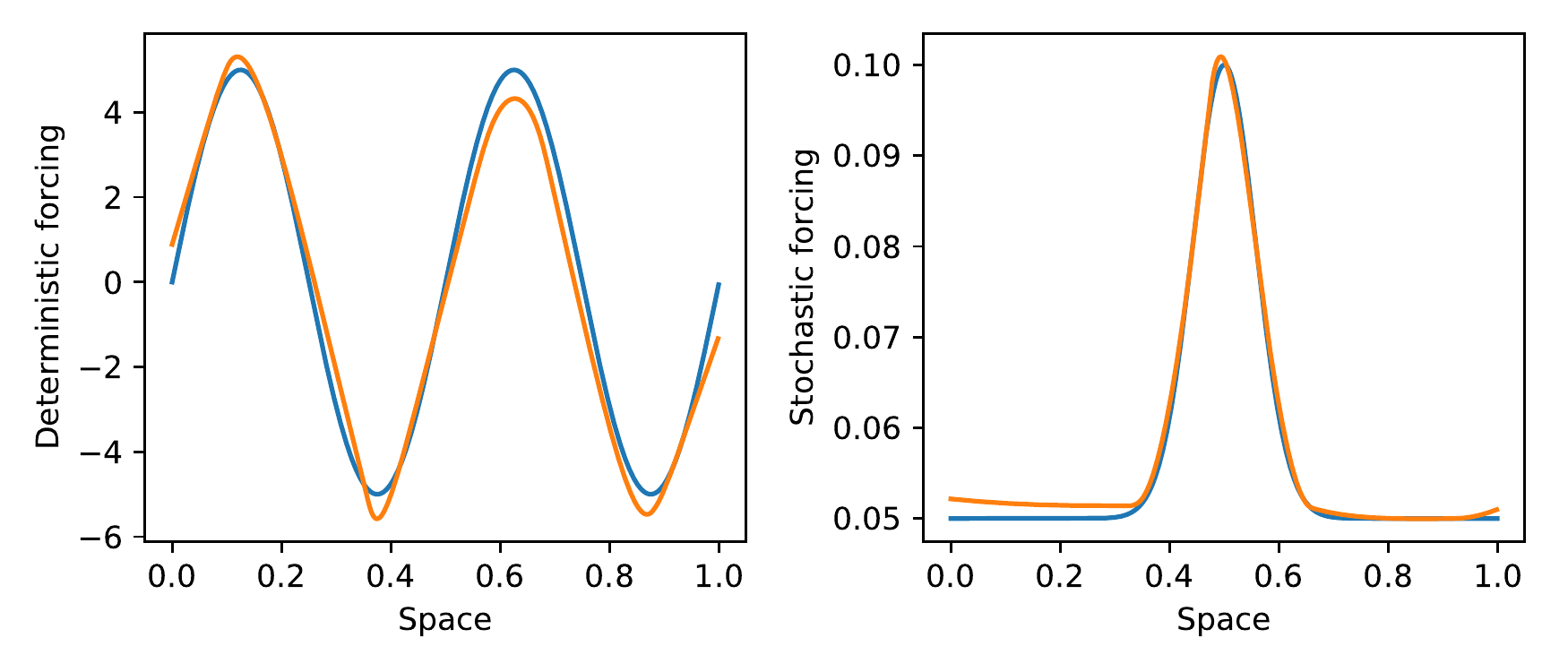}
    \caption{\review{Training data from the stochastic wave equation, and space dependent deterministic and stochastic forcing functions learned through the reformulated numerical solution method.}}
    \label{fig:example 7 wave pde}
\end{figure}
\begin{figure}[ht]
    \centering
    \includegraphics[height=0.17\textheight]{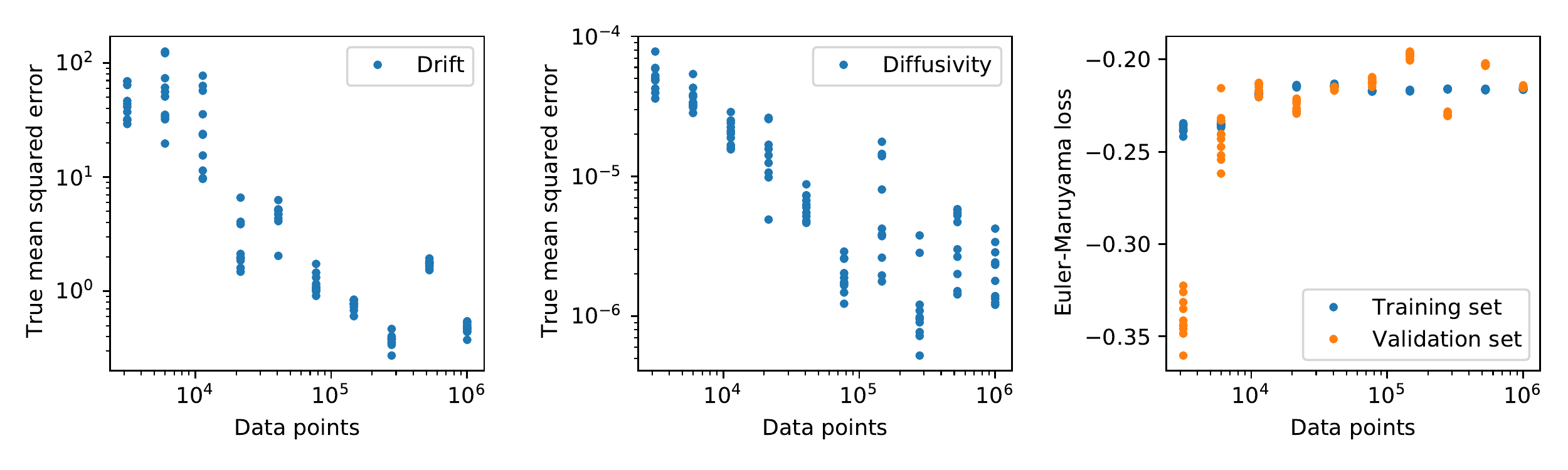}
    \caption{\review{Experimental results for convergence of the neural network approximations to the true forcing functions (drift and diffusivity) for increasing  number of available data points. Each data point in the panels indicates the result of an entire training process, with varying number of available data points but constant number of training updates.}}
    \label{fig:example 7 wave pde loss experiment}
\end{figure}
}

\section{Discussion}\label{sec:discussion}

Training neural networks with loss functions based on numerical integrators such as \eqref{eq:lossfunction EM}, \eqref{eq:lossfunction milstein}, but also \eqref{eq:lossfunction EM langevin}, has several limitations.
If the time step between many samples is too small, we cannot accurately identify the drift, because the diffusivity term will dominate. This could be mitigated by starting with small $\deltat$, estimating the diffusivity, and then estimating the drift with subsampled trajectories.
Even with infinite data the drift is difficult to estimate, because the time step also has to go to zero~\cite{Pavliotis-2014}.
Conversely, if the time step is too big, we cannot accurately identify the diffusivity.
If the dynamics of the coarse-grained observables include rare events, then learning the corresponding SDE is a challenge, because these events will not be present in a lot of snapshots, and hence not a lot of data is available to learn them. Using a loss function based on  L\'{e}vy noise SDE integrators~\cite{fang-2022} might provide a useful surrogate for SDEs with rare-events.



There are many possible extensions and applications of our work.
An important task is to find explicit probability density functions for integrators using other noise types, such as L\'{e}vy or Poisson noise~\cite{fang-2022}\review{, or integrators of higher order~\cite{roberts-2012a}. Integrating additional noise from measurement devices on top of the inherent stochasticity of the dynamics will also be important.}
Analyzing integrators based on other types of calculus such as Heun's~\cite{Burrage-2004} method may help impose more variegated types of priors during training.
It is a predictor-corrector method,
$\tilde{\X}_1=\X_0+\drift(\X_0)\deltat+\std(\X_0)z; \X_1=\X_0+1/2[\drift(\X_0)+\drift(\tilde{\X})]\deltat+1/2[\std(\X_0)+\std(\tilde{\X}_1)]z;  z\sim\normal(0,{\deltat}),$ based on Stratonovich calculus, and will give rise to implicit neural networks~\cite{Rico-Martinez-1995,anderson-1996}.
Using the adjoint method~\cite{liu-2020} allows us to propagate the loss gradients back through longer time series.
The choice of numerical approximation (\ref{eq:general time stepping}) entails a systematic error for the data generating distribution which can impact adversely on the capability to generalize beyond the training data. This effect can be minimized by higher-order numerical methods and analysed through modified equation analysis~\cite{Zygalakis-2011}.

In terms of applications, many more particle- and agent-based models can be coarse grained.
For many of these, it is important to find coarse variables, and latent space techniques such as VAE-type architectures and Diffusion Maps discussed in \secref{sec:latent spaces} will help to identify them.
These techniques can help to construct local SDE models in the process of larger-scale simulations to guide sampling and exploration.
Another clear next step based on numerical analysis is the identification of SPDEs, \review{for which we have demonstrated how our approach can be used to approximate forcing terms. Identification of entire SPDEs from data, including all terms, is still an open issue.}

\begin{acknowledgments}
\review{The authors wish to dedicate this manuscript to the memory of Prof. Alexei Makeev, a key contributor to the work, a friend, an inspiring and knowledgeable collaborator, whom we lost to Covid last year. 

We thank A.~J. Roberts for insightful discussions.}
The research of F.D. has been partially funded by the Deutsche Forschungsgemeinschaft (DFG) - Project-ID 468830823.
The research of S.R. has been partially funded by the Deutsche Forschungsgemeinschaft (DFG) - Project-ID 318763901 - SFB1294.
The work of I.G.K, N.E., and T.B. was partially supported by the US Department of Energy and the US Air Force Office of Scientific Research.
\end{acknowledgments}

\bibliography{mainNotes}

\begin{thebibliography}{10}

\bibitem{tensorflow2015-whitepaper}
M.~Abadi, A.~Agarwal, P.~Barham, E.~Brevdo, Z.~Chen, C.~Citro, G.~S. Corrado,
  A.~Davis, J.~Dean, M.~Devin, S.~Ghemawat, I.~Goodfellow, A.~Harp, G.~Irving,
  M.~Isard, Y.~Jia, R.~Jozefowicz, L.~Kaiser, M.~Kudlur, J.~Levenberg,
  D.~Man\'{e}, R.~Monga, S.~Moore, D.~Murray, C.~Olah, M.~Schuster, J.~Shlens,
  B.~Steiner, I.~Sutskever, K.~Talwar, P.~Tucker, V.~Vanhoucke, V.~Vasudevan,
  F.~Vi\'{e}gas, O.~Vinyals, P.~Warden, M.~Wattenberg, M.~Wicke, Y.~Yu, and
  X.~Zheng.
\newblock {TensorFlow}: Large-scale machine learning on heterogeneous systems,
  2015.
\newblock Software available from tensorflow.org.

\bibitem{allen-2017}
L.~J. Allen.
\newblock A primer on stochastic epidemic models: Formulation, numerical
  simulation, and analysis.
\newblock {\em Infectious Disease Modelling}, 2(2):128--142, May 2017.

\bibitem{anderson-1996}
J.~Anderson, I.~Kevrekidis, and R.~{Rico-Martinez}.
\newblock A comparison of recurrent training algorithms for time series
  analysis and system identification.
\newblock {\em Computers \& Chemical Engineering}, 20:S751--S756, Jan. 1996.

\bibitem{Arbabi-2019}
H.~Arbabi and T.~Sapsis.
\newblock Generative stochastic modeling of strongly nonlinear flows with
  non-gaussian statistics.
\newblock {\em arXiv:1908.08941}, 2019.

\bibitem{berry-2013}
T.~Berry, J.~R. Cressman, Z.~Greguri\'{c}-Feren\^{c}ek, and T.~Sauer.
\newblock Time-scale separation from diffusion-mapped delay coordinates.
\newblock {\em SIAM Journal on Applied Dynamical Systems}, 12(2):618--649, Jan.
  2013.

\bibitem{berry-2015b}
T.~Berry, D.~Giannakis, and J.~Harlim.
\newblock Nonparametric forecasting of low-dimensional dynamical systems.
\newblock {\em Physival Review E}, 91(3), 3 2015.

\bibitem{bertalan-2019}
T.~Bertalan, F.~Dietrich, I.~Mezi{\'{c}}, and I.~G. Kevrekidis.
\newblock On learning hamiltonian systems from data.
\newblock {\em Chaos: An Interdisciplinary Journal of Nonlinear Science},
  29(12):121107, Dec. 2019.

\bibitem{bongard-2007}
J.~Bongard and H.~Lipson.
\newblock Automated reverse engineering of nonlinear dynamical systems.
\newblock {\em Proceedings of the National Academy of Sciences},
  104(24):9943--9948, 2007.

\bibitem{Bortz-1975}
A.~B. Bortz, M.~H. Kalos, and J.~L. Lebowitz.
\newblock A new algorithm for {Monte Carlo} simulation of {Ising} spin systems.
\newblock {\em Journal of Computational Physics}, 17(1):10--18, 1975.

\bibitem{brunton-2016c}
S.~L. Brunton, J.~L. Proctor, and J.~N. Kutz.
\newblock Discovering governing equations from data by sparse identification of
  nonlinear dynamical systems.
\newblock {\em Proceedings of the National Academy of Sciences of the United
  States of America}, 113(15):3932--3937, 2016.

\bibitem{Burrage-2004}
K.~Burrage, P.~M. Burrage, and T.~Tian.
\newblock Numerical methods for strong solutions of stochastic differential
  equations: an overview.
\newblock {\em Proceedings of the Royal Society of London. Series A:
  Mathematical, Physical and Engineering Sciences}, 460(2041):373--402, 2004.

\bibitem{chen-2018}
R.~T.~Q. Chen, Y.~Rubanova, J.~Bettencourt, and D.~Duvenaud.
\newblock Neural ordinary differential equations.
\newblock In {\em NeurIPS conference 2018}, 2018.

\bibitem{coifman-2006}
R.~R. Coifman and S.~Lafon.
\newblock Diffusion maps.
\newblock {\em Applied and Computational Harmonic Analysis}, 21(1):5--30, July
  2006.

\bibitem{crandall-1992}
M.~G. Crandall, H.~Ishii, and P.-L. Lions.
\newblock User's guide to viscosity solutions of second order partial
  differential equations.
\newblock {\em Bull. Amer. Math. Soc.}, 27:1--67, 1992.

\bibitem{crandall-1983}
M.~G. Crandall and P.-L. Lions.
\newblock Viscosity solutions of hamilton-jacobi equations.
\newblock {\em Transactions of the American Mathematical Society},
  277(1):1--42, 1983.

\bibitem{ditlevsen-2005}
S.~Ditlevsen and P.~Lansky.
\newblock Estimation of the input parameters in the {{Ornstein-Uhlenbeck}}
  neuronal model.
\newblock {\em Physical Review E}, 71(1):011907, Jan. 2005.

\bibitem{dsilva-2018}
C.~J. Dsilva, R.~Talmon, R.~R. Coifman, and I.~G. Kevrekidis.
\newblock Parsimonious representation of nonlinear dynamical systems through
  manifold learning: A chemotaxis case study.
\newblock {\em Applied and Computational Harmonic Analysis}, 44(3):759--773,
  May 2018.

\bibitem{fang-2022}
C.~Fang, Y.~Lu, T.~Gao, and J.~Duan.
\newblock An end-to-end deep learning approach for extracting stochastic
  dynamical systems with \$\textbackslash alpha\$-stable {{L}}\textbackslash
  'evy noise.
\newblock {\em arXiv:2201.13114 [cs, stat]}, Feb. 2022.

\bibitem{geneva-2022}
N.~Geneva and N.~Zabaras.
\newblock Transformers for modeling physical systems.
\newblock {\em Neural Networks}, 146:272--289, Feb. 2022.

\bibitem{Gillespie-1976}
D.~T. Gillespie.
\newblock A general method for numerically simulating the stochastic time
  evolution of coupled chemical reactions.
\newblock {\em Journal of Computational Physics}, 22(4):403--434, 1976.

\bibitem{Gillespie-2000}
D.~T. Gillespie.
\newblock The chemical langevin equation.
\newblock {\em The Journal of Chemical Physics}, 113(1):297--306, July 2000.

\bibitem{gonzalez-1998}
R.~Gonz{\'{a}}lez-Garc{\'{\i}}a, R.~Rico-Mart{\'{\i}}nez, and I.~Kevrekidis.
\newblock Identification of distributed parameter systems: A neural net based
  approach.
\newblock {\em Computers {\&} Chemical Engineering}, 22:S965--S968, Mar. 1998.

\bibitem{goodfellow-2014}
I.~Goodfellow, J.~Pouget-Abadie, M.~Mirza, B.~Xu, D.~Warde-Farley, S.~Ozair,
  A.~Courville, and Y.~Bengio.
\newblock Generative adversarial nets.
\newblock In Z.~Ghahramani, M.~Welling, C.~Cortes, N.~D. Lawrence, and K.~Q.
  Weinberger, editors, {\em Advances in Neural Information Processing Systems
  27}, pages 2672--2680. Curran Associates, Inc., 2014.

\bibitem{Graves-2013}
A.~Graves.
\newblock Generating sequences with recurrent neural networks.
\newblock {\em arXiv:1308.0850}, 2013.

\bibitem{Greydanus-2019}
S.~Greydanus, M.~Dzamba, and J.~Yosinski.
\newblock Hamiltonian neural networks.
\newblock In {\em Proceedings of the 33rd International Conference on Neural
  Information Processing Systems}, 2019.

\bibitem{hasan-2022}
A.~Hasan, J.~M.~Pereira, S.~Farsiu, and V.~Tarokh.
\newblock Identifying {{Latent Stochastic Differential Equations}}.
\newblock {\em IEEE Transactions on Signal Processing}, 70:89--104, 2022.

\bibitem{he-2016}
K.~He, X.~Zhang, S.~Ren, and J.~Sun.
\newblock Deep residual learning for image recognition.
\newblock In {\em 2016 {IEEE} Conference on Computer Vision and Pattern
  Recognition ({CVPR})}. {IEEE}, June 2016.

\bibitem{Higham-2001}
D.~J. Higham.
\newblock An algorithmic introduction to numerical simulation of stochastic
  differential equations.
\newblock {\em {SIAM} Review}, 43(3):525--546, Jan. 2001.

\bibitem{jalal-2017}
A.~Jalal, A.~Ilyas, C.~Daskalakis, and A.~G. Dimakis.
\newblock The robust manifold defense: Adversarial training using generative
  models.
\newblock {\em arXiv:1712.09196v5}, 2017.

\bibitem{Jin-2020c}
P.~Jin, Z.~Zhang, I.~G. Kevrekidis, and G.~E. Karniadakis.
\newblock Learning poisson systems and trajectories of autonomous systems via
  poisson neural networks.
\newblock {\em arXiv:2012.03133}, 2020.

\bibitem{karatzas-1998}
I.~Karatzas and S.~E. Shreve.
\newblock {\em Brownian Motion and Stochastic Calculus}.
\newblock Springer New York, 1998.

\bibitem{kermack-1927}
W.~O. Kermack and A.~G. McKendrick.
\newblock A contribution to the mathematical theory of epidemics.
\newblock {\em Proceedings of the Royal Society of London. Series A, Containing
  Papers of a Mathematical and Physical Character}, 115(772):700--721, Aug.
  1927.

\bibitem{kidger-2021}
P.~Kidger.
\newblock On {{Neural Differential Equations}}.
\newblock Master's thesis, University of Oxford, 2021.

\bibitem{kidger-2021a}
P.~Kidger, J.~Foster, X.~Li, and T.~J. Lyons.
\newblock Neural {{SDEs}} as {{Infinite-Dimensional GANs}}.
\newblock In {\em Proceedings of the 38th {{International Conference}} on
  {{Machine Learning}}}, pages 5453--5463. {PMLR}, July 2021.

\bibitem{kingma-2013}
D.~P. Kingma and M.~Welling.
\newblock Auto-encoding variational bayes.
\newblock In {\em Proceedings of the 2nd International Conference on Learning
  Representations (ICLR)}, Dec. 2014.

\bibitem{klus-2020}
S.~Klus, F.~N\"{u}ske, S.~Peitz, J.-H. Niemann, C.~Clementi, and
  C.~Sch\"{u}tte.
\newblock Data-driven approximation of the koopman generator: Model reduction,
  system identification, and control.
\newblock {\em Physica D: Nonlinear Phenomena}, 406:132416, May 2020.

\bibitem{Kobyzev-2019}
I.~Kobyzev, S.~J.~D. Prince, and M.~A. Brubaker.
\newblock {Normalizing Flows: An Introduction and Review of Current Methods}.
\newblock {\em IEEE Transactions on Pattern Analysis and Machine Intelligence},
  Aug. 2019.

\bibitem{lee-2017c}
J.~Lee, J.~Sohl-dickstein, J.~Pennington, R.~Novak, S.~Schoenholz, and
  Y.~Bahri.
\newblock Deep neural networks as gaussian processes.
\newblock In {\em International Conference on Learning Representations}, 2018.

\bibitem{lehmberg-2020}
D.~Lehmberg, F.~Dietrich, G.~Köster, and H.-J. Bungartz.
\newblock datafold: data-driven models for point clouds and time series on
  manifolds.
\newblock {\em Journal of Open Source Software}, 5(51):2283, July 2020.

\bibitem{li2020scalable}
X.~Li, T.-K.~L. Wong, R.~T. Chen, and D.~Duvenaud.
\newblock Scalable gradients for stochastic differential equations.
\newblock {\em arXiv:2001.01328}, 2020.

\bibitem{liu-2020}
J.~Liu, Z.~Long, R.~Wang, J.~Sun, and B.~Dong.
\newblock Rode-net: Learning ordinary differential equations with randomness
  from data.
\newblock {\em arXiv:2006.02377}, 2020.

\bibitem{Lu-2021}
L.~Lu, P.~Jin, G.~Pang, Z.~Zhang, and G.~E. Karniadakis.
\newblock Learning nonlinear operators via {DeepONet} based on the universal
  approximation theorem of operators.
\newblock {\em Nature Machine Intelligence}, 3(3):218--229, 3 2021.

\bibitem{makeev-2002}
A.~G. Makeev, D.~Maroudas, and I.~G. Kevrekidis.
\newblock Coarse stability and bifurcation analysis using stochastic
  simulators: {Kinetic Monte Carlo} examples.
\newblock {\em The Journal of Chemical Physics}, 116(23):10083--10091, June
  2002.

\bibitem{makeev-2002b}
A.~G. Makeev, D.~Maroudas, A.~Z. Panagiotopoulos, and I.~G. Kevrekidis.
\newblock Coarse bifurcation analysis of kinetic monte carlo simulations: A
  lattice-gas model with lateral interactions.
\newblock {\em The Journal of Chemical Physics}, 117(18):8229--8240, Nov. 2002.

\bibitem{Makeev-2017}
A.~G. Makeev and N.~L. Semendyaeva.
\newblock A basic lattice model of an excitable medium: Kinetic monte carlo
  simulations.
\newblock {\em Mathematical Models and Computer Simulations}, 9(5):636--648,
  2017-09.

\bibitem{mannella-2006}
R.~Mannella.
\newblock Numerical {{Stochastic Integration}} for {{Quasi-Symplectic Flows}}.
\newblock {\em SIAM Journal on Scientific Computing}, 27(6):2121--2139, Jan.
  2006.

\bibitem{mezic-2005}
I.~Mezi{\'{c}}.
\newblock Spectral properties of dynamical systems, model reduction and
  decompositions.
\newblock {\em Nonlinear Dynamics}, 41(1):309--325, Aug. 2005.

\bibitem{Milshtejn-1975}
G.~N. Mil'shtejn.
\newblock Approximate integration of stochastic differential equations.
\newblock {\em Theory of Probability {\&} Its Applications}, 19(3):557--562,
  June 1975.

\bibitem{morrill-2021}
J.~Morrill, C.~Salvi, P.~Kidger, and J.~Foster.
\newblock Neural {{Rough Differential Equations}} for {{Long Time Series}}.
\newblock In {\em Proceedings of the 38th {{International Conference}} on
  {{Machine Learning}}}, pages 7829--7838. {PMLR}, July 2021.

\bibitem{Pavliotis-2014}
G.~A. Pavliotis.
\newblock {\em Stochastic Processes and Applications}.
\newblock Springer New York, 2014.

\bibitem{perez2010stochastic}
R.~Perez-Carrasco and J.~Sancho.
\newblock Stochastic algorithms for discontinuous multiplicative white noise.
\newblock {\em Physical Review E}, 81(3):032104, 2010.

\bibitem{rahimi-2006}
A.~Rahimi and B.~Recht.
\newblock Unsupervised regression with applications to nonlinear system
  identification.
\newblock In {\em Proceedings of the 19th International Conference on Neural
  Information Processing Systems}, NIPS'06, pages 1113--1120, Cambridge, MA,
  USA, 2006. MIT Press.

\bibitem{rico-martinez-1994}
R.~Rico-Martinez, J.~Anderson, and I.~Kevrekidis.
\newblock Continuous-time nonlinear signal processing: a neural network based
  approach for gray box identification.
\newblock In {\em Proceedings of {IEEE} Workshop on Neural Networks for Signal
  Processing}. {IEEE}, 1994.

\bibitem{Rico-Martinez-1995}
R.~Rico-Mart{\'{i}}nez and I.~Kevrekidis.
\newblock {\em Nonlinear system identification using neural networks: dynamics
  and instabilities}, chapter~16, pages 409--442.
\newblock Elsevier Science, 1995.

\bibitem{Rico-Martinez-1992}
R.~Rico-Mart{\'{i}}nez, K.~Krischer, I.~Kevrekidis, M.~Kube, and J.~Hudson.
\newblock {Discrete}- vs. {continuous}-{time} {nonlinear} {signal} {processing}
  {of} {Cu} {electrodissolution} {data}.
\newblock {\em Chemical Engineering Communications}, 118(1):25--48, Nov. 1992.

\bibitem{roberts-2012a}
A.~J. Roberts.
\newblock Modify the {{Improved Euler}} scheme to integrate stochastic
  differential equations.
\newblock {\em arXiv:1210.0933 [math]}, Oct. 2012.

\bibitem{salvi-2021}
C.~Salvi, M.~Lemercier, and A.~Gerasimovics.
\newblock Neural {{Stochastic Partial Differential Equations}}:
  {{Resolution-Invariant Learning}} of {{Continuous Spatiotemporal Dynamics}}.
\newblock {\em arXiv:2110.10249 [cs]}, Oct. 2021.

\bibitem{schoelkopf-2018}
B.~Sch\"{o}lkopf and A.~J. Smola.
\newblock {\em Learning with Kernels: Support Vector Machines, Regularization,
  Optimization, and Beyond}.
\newblock The MIT Press, 2018.

\bibitem{Siettos-2013}
C.~Siettos and L.~Russo.
\newblock Mathematical modeling of infectious disease dynamics.
\newblock {\em Virulence}, 4:295 -- 306, 2013.

\bibitem{Song-2021}
Y.~Song, J.~Sohl-Dickstein, D.~P. Kingma, A.~Kumar, S.~Ermon, and B.~Poole.
\newblock Score-based generative modeling through stochastic differential
  equations.
\newblock In {\em International Conference on Learning Representations}, 2021.

\bibitem{walsh-2006}
J.~B. Walsh.
\newblock On numerical solutions of the stochastic wave equation.
\newblock {\em Illinois Journal of Mathematics}, 50(1-4):991--1018, Jan. 2006.

\bibitem{Yang-2020b}
L.~Yang, C.~Daskalakis, and G.~E. Karniadakis.
\newblock Generative ensemble-regression: Learning stochastic dynamics from
  discrete particle ensemble observations.
\newblock {\em arXiv:2008.01915v1}, 2020.

\bibitem{Yang-2021}
S.~Yang, S.~W.~K. Wong, and S.~C. Kou.
\newblock Inference of dynamic systems from noisy and sparse data via
  manifold-constrained gaussian processes.
\newblock {\em Proceedings of the National Academy of Sciences},
  118(15):e2020397118, 2021.

\bibitem{yildiz-2018}
C.~Yildiz, M.~Heinonen, J.~Intosalmi, H.~Mannerstrom, and H.~Lahdesmaki.
\newblock Learning stochastic differential equations with with {Gaussian}
  {Processes} without gradient matching.
\newblock In {\em 2018 {IEEE} 28th International Workshop on Machine Learning
  for Signal Processing ({MLSP})}. {IEEE}, Sept. 2018.

\bibitem{zhu-2020}
A.~Zhu, P.~Jin, and Y.~Tang.
\newblock Deep hamiltonian networks based on symplectic integrators.
\newblock {\em arXiv:2004.13830}, 2020.

\bibitem{zhu-2022}
Y.~Zhu, Y.-H. Tang, and C.~Kim.
\newblock Learning {{Stochastic Dynamics}} with {{Statistics-Informed Neural
  Network}}, Feb. 2022.

\bibitem{Zygalakis-2011}
K.~C. Zygalakis.
\newblock On the existence and the applications of modified equations for
  stochastic differential equations.
\newblock {\em {SIAM} Journal on Scientific Computing}, 33(1):102--130, Jan.
  2011.

\end{thebibliography}


\newpage

\appendix




\section{Instructions for the software}

The code to reproduce the results of the computational experiments is available at \url{https://gitlab.com/felix.dietrich/sde-identification}.
Tab.~\ref{tab:software req} lists the software we used to test if the code can be compiled and actually produces the experimental results.
\begin{table}[ht]
    \centering
    \caption{\label{tab:software req}Software we used to test that the code in the supplement can be compiled and run.}
    \begin{tabular}{ll}
    \toprule
       Software  & Version \\
         \midrule
       Microsoft Windows & 10.0.19042 \\
       Python & 3.8 \\
        Microsoft Visual Studio Community 2019 & 16.9.0\\
        Microsoft Visual C++ & 2019 \\
         \bottomrule
    \end{tabular}
\end{table}

All required python packages can be installed by running
``\texttt{pip install -r requirements.txt}'' in the folder where the \texttt{requirements.txt} file is placed.
The toy examples and lattice computations are all available in separate files, the mapping is listed in Tab.~\ref{tab:notebooks}. 
To compile the \texttt{kMCLattice.dll}, the Microsoft Visual Studio solution file in \texttt{kmc/kMCLattice\_dll} can be used. Make sure to use an \texttt{x64} setup (for both Python and the \texttt{.dll} file). The file needs to be placed in the same folder as the Python script \texttt{kmc/KMC\_SIRS3.py}.
The SSA results can be obtained by setting both \texttt{Dif1} and \texttt{Dif2} to $1000$ in the file \texttt{kmc/KMC\_SIRS3.py}.

\begin{table}[ht]
    \centering
    \caption{\label{tab:notebooks}Mapping of computational experiments to Jupyter notebooks (.ipynb) / python (.py) files.}
    \begin{tabular}{ll}
    \toprule
       Experiment  & File \\
         \midrule
       Toy example: 1D, cubic drift, linear diff. (E.-M.) & example3 - 1d sde-cubic.ipynb \\
       Toy example: 1D, cubic drift, linear diff. (Milstein) & example3 - 1d sde-cubic-with-milstein.ipynb \\
       Toy example: 3D, linear drift, lower-triangular diff. & example4 - 3d sde-nondiagonal.ipynb \\
       Toy example: 3D, linear drift, SPD diff. & example4 - 3d sde-spd.ipynb \\
       Toy example: nD (1-19D), cubic drift, linear diff. & example3 - nd sde-cubic.ipynb \\
       Kinetic Monte-Carlo simulations & kmc/KMC\_SIRS3.py \\
       Stochastic wave equation & example7 - Wave SPDE.ipynb \\
       Langevin dynamics & example8 - nonGaussian.ipynb \\
         \bottomrule
    \end{tabular}
\end{table}

\section{Additional information regarding the computational experiments}

The following subsections contain additional descriptions and results for the computational experiments. They are not necessary to follow the main results described in the paper.

\subsection{Milstein loss}

\begin{figure}[ht]
\centering
\includegraphics[width=0.8\textwidth]{figures/density_milstein_vs_euler}
\includegraphics[width=0.8\textwidth]{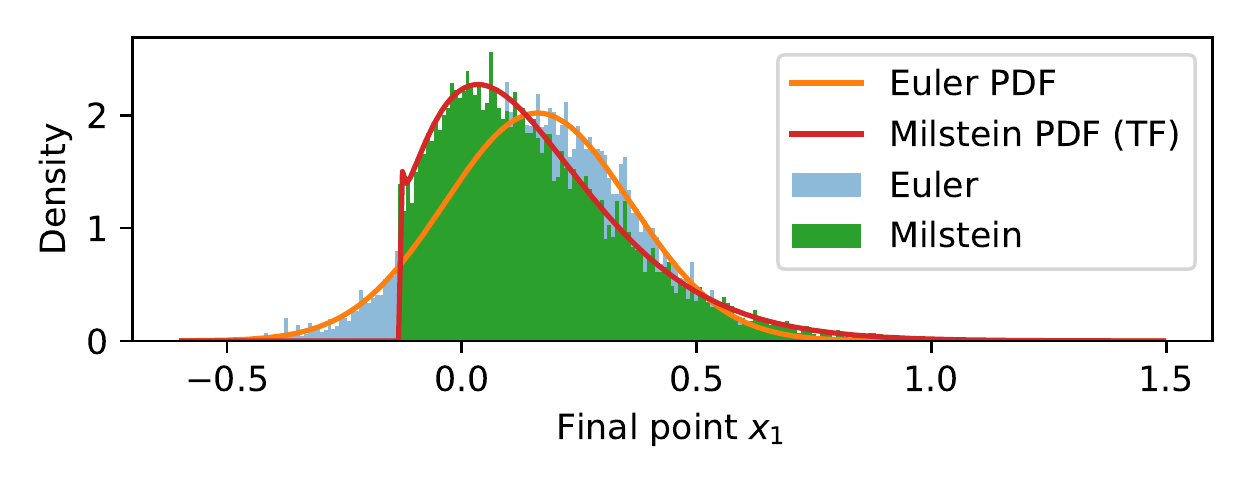}
\caption{\label{supp:fig:euler vs milstein pdf appendix}Probability density functions and sampled data from the two numerical integrator templates. The top panel shows the PDFs with the correct Bessel function with real-valued parameter $\nu=-1/2$, using \texttt{scipy.stats.ncx2} for the PDF of the non-centerd $\chi^2$ distribution. The bottom panel shows the PDF with an integer valued parameter $\nu=0$ (the only one available in TensorFlow).}
\end{figure}

The TensorFlow implementation for the Milstein loss uses a slightly different Bessel function, with integer-valued parameter $\nu=0$ instead of the required $\nu=-1/2$ (no other Bessel function was available). Figure~\ref{supp:fig:euler vs milstein pdf appendix} demonstrates the difference is quite small.

\subsection{Toy example: cubic drift, linear diffusivity}
\begin{figure}[ht]
    \centering
    \includegraphics[width=1\textwidth]{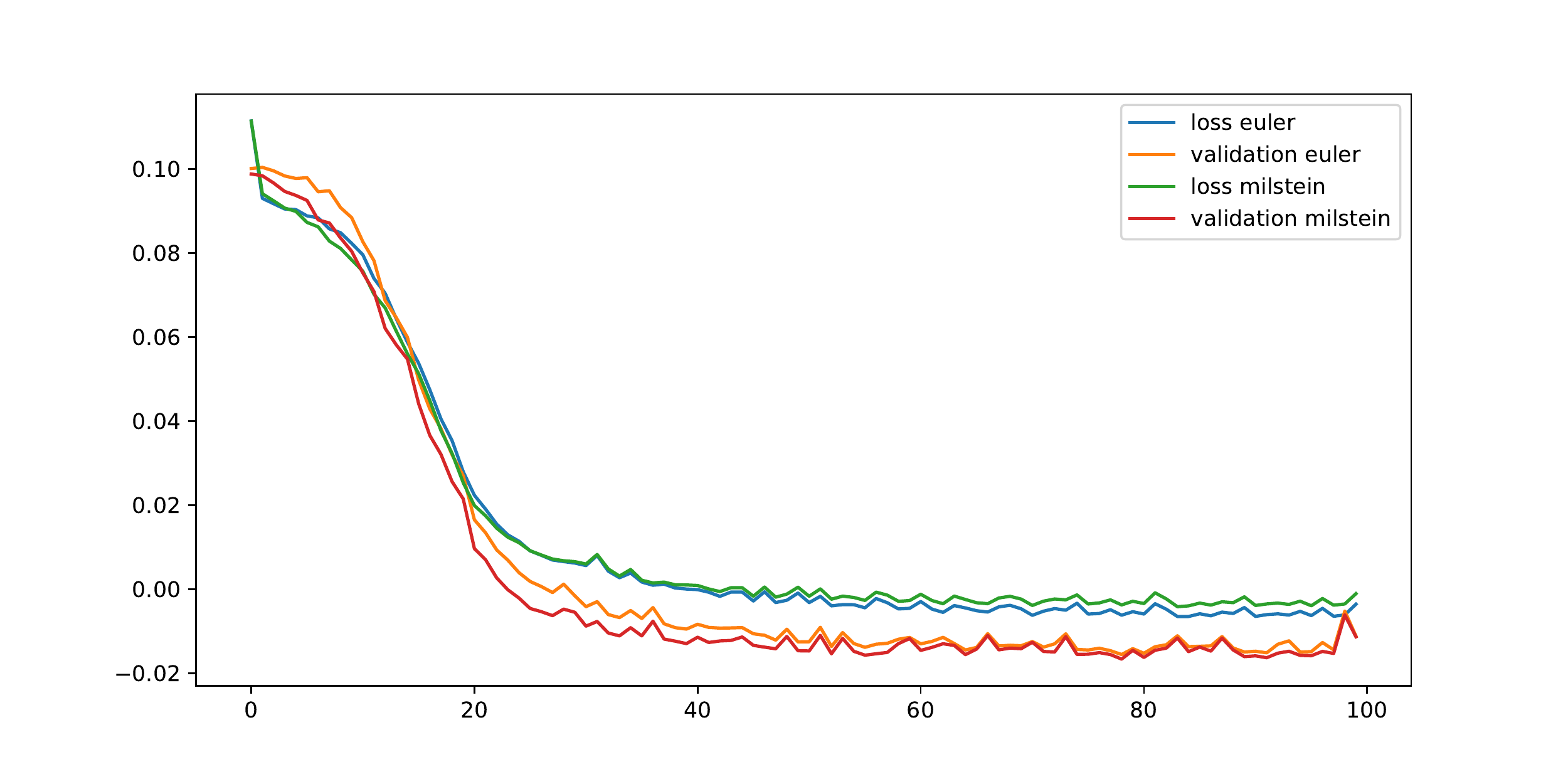}
    \caption{Milstein vs. Euler-Maruyama loss curves.}
    \label{supp:fig:euler vs milstein loss appendix}
\end{figure}

The loss and validation loss curves for Euler-Maruyama and Milstein schemes are shown in figure~\ref{supp:fig:euler vs milstein loss appendix}.

\subsection{Toy example: non-diagonal diffusivity}\label{sec:example nondiagonal}
\begin{figure}[ht]
    \centering
    \includegraphics[width=1\textwidth]{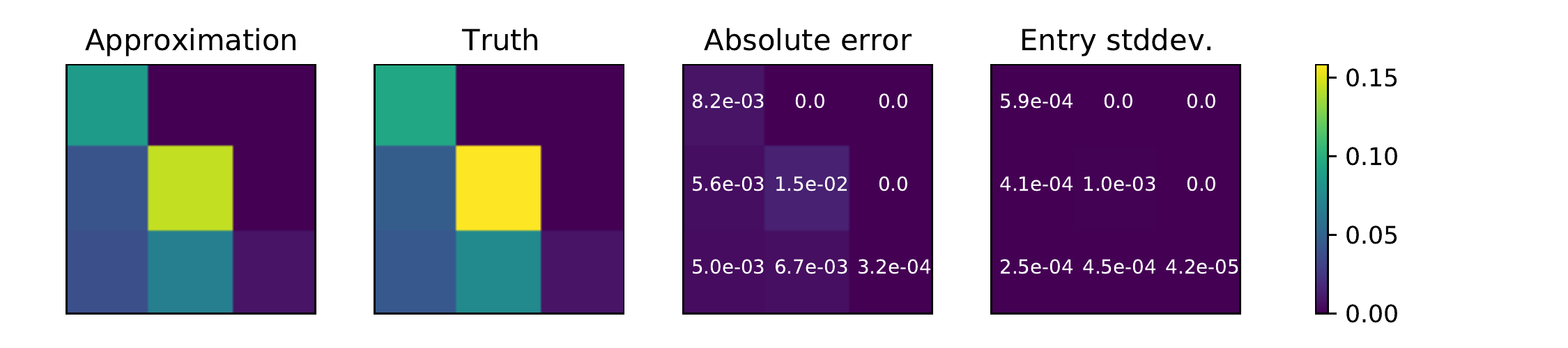}
    \includegraphics[width=1\textwidth]{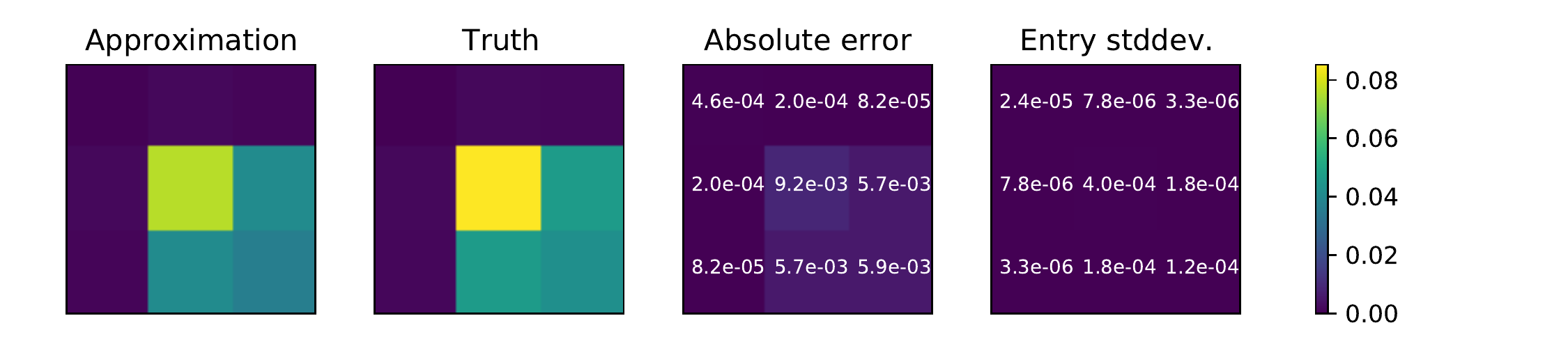}
    \caption{\label{fig:example_3dnondiag_matrixcompare} Lower-triangular (top row) and full symmetric, positive definite (bottom row) diffusivity matrix approximation. The panels show the averaged network output over the uniformly sampled data inside the training region, the true diffusivity matrices, the absolute error between the results shown in the first and second columns, and the deviation of the network matrix output over the training region. The color range is the same for all four plots.}
\end{figure}

We now demonstrate that our approach can learn non-diagonal diffusivity matrices. We sample initial points $\boldsymbol{\X}_0\in[-.3,.3]^3$, randomly sample a lower-triagonal diffusivity matrix $\boldsymbol{\Sigma}$ with positive eigenvalues, which we use as the (constant) diffusivity $\boldsymbol{\std}(\boldsymbol{\X})=\boldsymbol{\Sigma}$, and set the drift to be $\boldsymbol{f}(\boldsymbol{\X})=-\boldsymbol{\X}$. The absolute error between original and (the average over) identified matrix entries is smaller than $0.01$, and the standard deviation of the diffusivity values over the entire data set is smaller than $0.003$ for all elements of the matrix.
Note that the network $\stdNN$ is still a nonlinear function over the state space, not a constant. To ensure the matrix has positive eigenvalues, we pass the real-valued output of the neurons that encode the diagonal of the matrix through a soft-plus activation function.
Figure~\ref{fig:example_3dnondiag_matrixcompare} shows that we learn a good approximation for $\Sigma$.

\subsection{Toy example: increasing dimension from 1D to 19D}

\begin{figure}[ht]
    \centering
    \includegraphics[width=.6\textwidth]{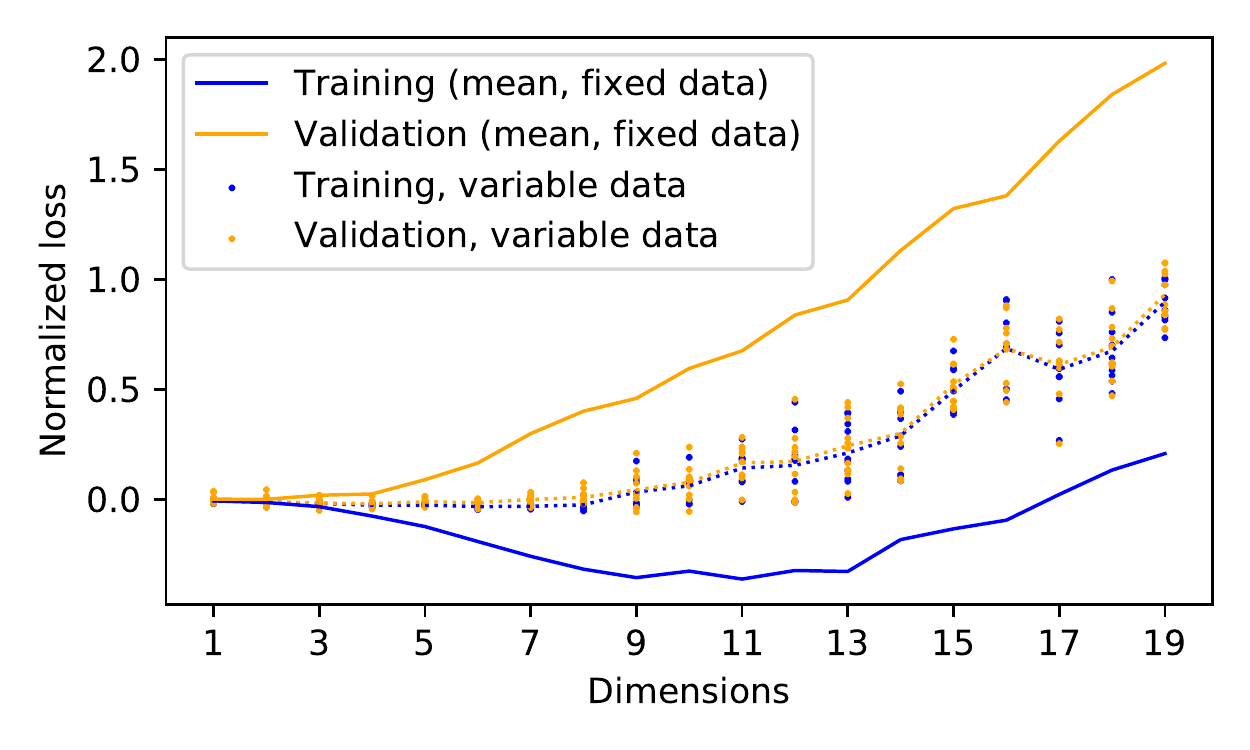}
    \caption{
    \label{fig:nd training}\review{Normalized training and validation loss when increasing the intrinsic dimension of the problem from $\ndim=1$ to $\ndim=19$. The solid lines show the results of keeping the number of samples constant at $\numpoints=10,000$, the dashed lines  show the average of what happens when increasing $\numpoints=10,000\times\ndim$, but training for the same number of iterations per dimension ($\text{Epochs}=1000/\ndim$).}}
\end{figure}

We define $\drift(\boldsymbol{x})=-(4\boldsymbol{x}^3 - 8\boldsymbol{x}+3)/2$ and $\std(\boldsymbol{x})=5e-2\boldsymbol{x}+0.5$, where all operations are meant coordinate-wise (e.g. $\boldsymbol{x}^3$ computes the third power of each individual coordinate of the vector). Figure~\ref{fig:nd training} (left) illustrates how the training and validation losses change when increasing $\ndim$ from $1$ to $19$. The loss ($\lossfunction_{\text{Euler-Maruyama}}$) is adapted to the changing dimensionality by adding $\log(2\pi)\ndim$, as described by \eqnref{eq:nd EM}. This normalization is necessary because (apparently) TensorFlow 2.4.1 does not add this constant when computing the log probability (See \url{https://www.tensorflow.org/probability/api_docs/python/tfp/distributions/MultivariateNormalDiag#log_prob}).
The increase in loss can be explained by the constant number of points ($\numpoints=10,000$) we used:
Increasing the intrinsic dimension of the problem by sampling the input data in the $\ndim$-dimensional cube $[-2,2]^\ndim$ causes the data sampling to get sparser and sparser.
By increasing the number of training data points linearly with the dimension (while keeping the number of training iterations per dimension constant), we can see that the training loss is relatively small even for $\ndim=12$ (\figref{fig:nd training}, right). \review{Additionally, the training loss tracks the validation loss much better, indicating better generalization properties of the network. This is reasonable, because the provided data is less sparse, even for higher dimensions.}

\subsection{Kinetic Monte-Carlo lattice simulations}

\begin{figure}[ht]
    \centering
    \includegraphics[width=.75\textwidth]{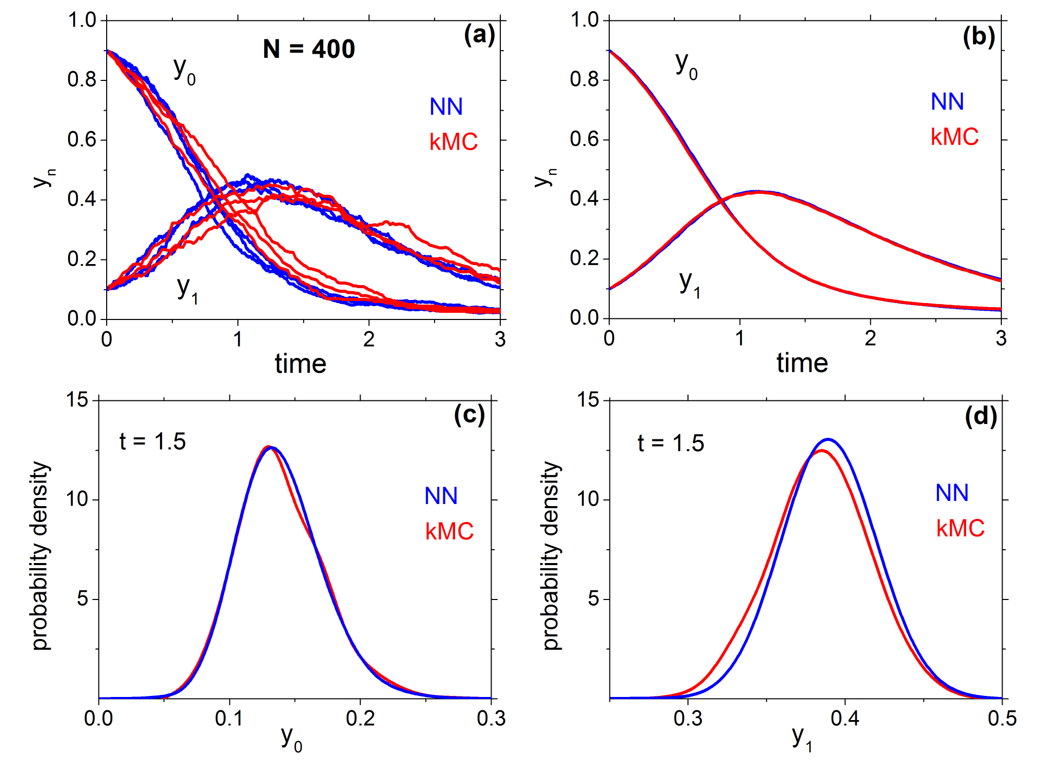}
    \caption{Blue (red) lines correspond to NN (kMC) results. (a) Three sample paths of $\SIRvar_0$ and $\SIRvar_1$; (b) Average of 200 paths; (c),(d) Probability density function for $\SIRvar_0$ and $\SIRvar_1$ at $t=1.5$ computed from $2\times 10^4$ sample paths. Parameters: $20\times 20$ lattice, $\migrationrate=50$; $N_{tr}=10000$, $t_{\text{max}}=0.05$, $\deltat=0.01$.}
    \label{fig:kmc_lattice_n400}
\end{figure}

\begin{figure}[ht]
    \centering
    \includegraphics[width=.33\textwidth]{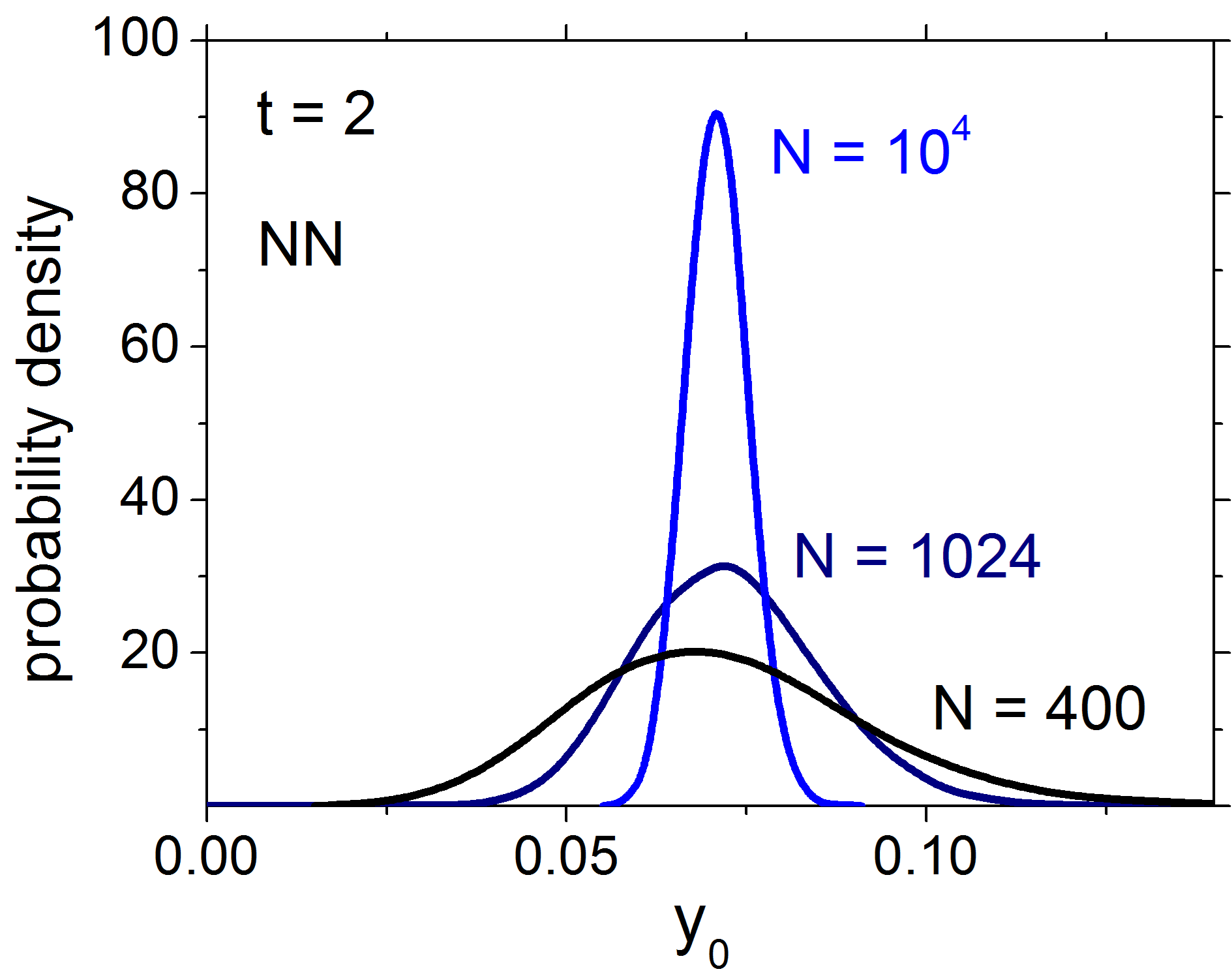}
    \caption{Influence of the system size on the neural network-estimated probability density function at $t=2$: $20\times 20$ (black), $32\times 32$ (navy), $100\times 100$ (blue).}
    \label{fig:kmc_lattice_systemsize}
\end{figure}

\begin{figure}[ht]
    \centering
    \includegraphics[width=.75\textwidth]{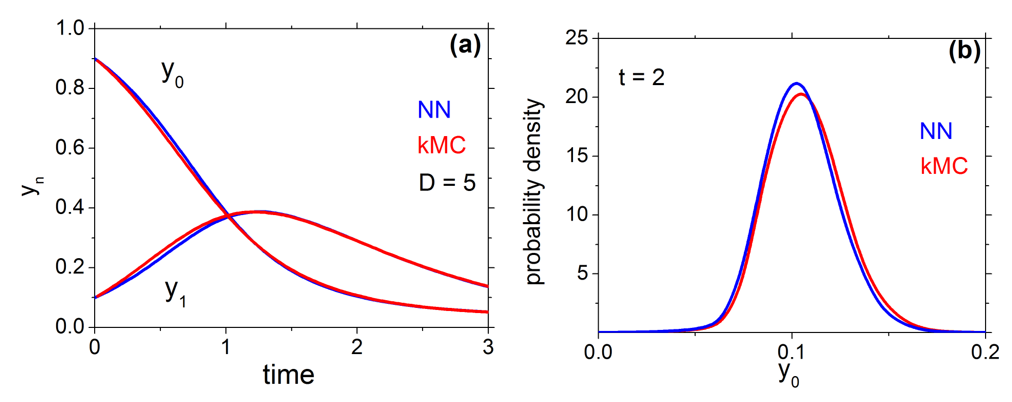}
    \caption{Relatively low migration rate. Parameters: $32\times 32$ lattice, $\migrationrate=5$; $N_{tr}=4000$, $t_{\text{max}}=0.4$, $\deltat=0.02$.}
    \label{fig:kmc_lattice_d5}
\end{figure}

Results on a smaller $20\times 20$ lattice are shown in \figref{fig:kmc_lattice_n400}.
\figref{fig:kmc_lattice_systemsize} shows that the algorithm works at different system sizes, even when $\numparticles$ is large---that is, even when the noise level is low. For $\numparticles\to\infty$, the system becomes deterministic and can be described by an ODE.
The data presented in \figref{fig:kmc_lattice_systemsize} correspond to a relatively fast migration rate $\migrationrate=\migrationrate_1=\migrationrate_2=50$. In this case, a well-mixed state is achieved and the kMC results are close to that obtained with the SSA algorithm. However, the network can also be trained to match the kMC results at lower migration rate values.
\figref{fig:kmc_lattice_d5} shows the simulation results for $\migrationrate=5$. Since the initial distribution on a lattice is random, we omit the first few time steps and select the data points for training only when a ``matured'' state is achieved. For the results presented in \figref{fig:kmc_lattice_d5}, the training process was started at $t_0=0.1$. 
At $\migrationrate=5$, the accuracy of the network is acceptable, but at lower $\migrationrate$ values, the coincidence of the network and kMC results decreased considerably.
This suggests that, around that value of $\migrationrate$ the dynamics may not effectively close in terms of the mean field variables, and that higher dimensional approximations (possibly including pair probabilities as additional independent variables) may become necessary. 

%
Similar results can be obtained on a smaller $20\times 20$ lattice.
In this case, as expected, the influence of the stochasticity becomes more pronounced.
In a separate experiment, we also confirmed the algorithm works at different system sizes, even when $\numparticles$ is large---that is, even when the noise level is low. For $\numparticles\to\infty$, the system becomes deterministic and can be described by an ODE.
At low migration rate values ($\migrationrate<5$), the approximation quality of the network decreased considerably as expected: The system can hardly be described in terms of two averaged concentrations because the species distribution on a lattice becomes inhomogeneous: Clustering occurs, and a higher dimensional SDE, possibly in terms of densities {\em as well as} pair probabilities, becomes necessary. 

\begin{figure}[ht]
    \centering
    \includegraphics[width=.75\textwidth]{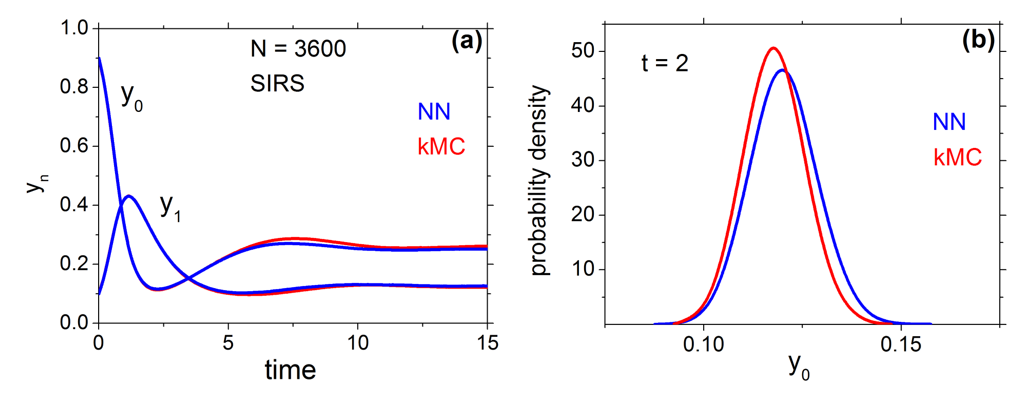}
    \caption{SIRS model with $k_3=0.2$, $60\times 60$ lattice, $\migrationrate=50$; $N_{tr}=5000$, $t_{\text{max}}=0.1$, $\deltat=0.01$. The left panel shows 200 averaged paths, the right panel shows the propagated densities until $t=2$.}
    \label{fig:kmc_lattice_d50}
\end{figure}

 When $k_3>0$, the system exhibits richer dynamic behavior. An example of simulations with $k_3=0.2$ is shown in \figref{fig:kmc_lattice_d50}.  The network model is capable of reproducing the correct steady state and damped oscillations (the mean field SIRS model has a focus-type steady state at these values of parameters).

\subsection{Learning latent spaces for the SIR model}

\begin{figure}[ht]
    \centering
    \includegraphics[width=.6\textwidth]{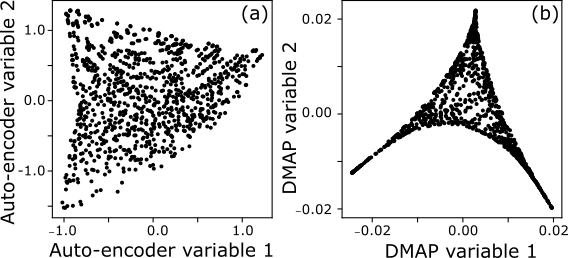}
    \caption{Latent spaces identified by SDE-integrator-informed auto-encoder (a), and with Diffusion Maps (b). Both show a triangular structure, as expected from a system with three intrinsic variables with a single conservation law ($y_0+y_1+y_2=\text{const}$).}
    \label{fig:damp_vs_autoencoder_latent}
\end{figure}

We identify latent spaces of six-dimensional data obtained from the kMC lattice simulations. \figref{fig:damp_vs_autoencoder_latent} illustrates that the latent spaces are both two-dimensional and triangular. This shape is due to the conservation law of the underlying model (when written in terms of fractions of the total population that are susceptible (S), infected (I), and removed (R), the law would be $S+I+R=1$).
\secref{sec:dmaps} outlines the Diffusion Maps algorithm, \secref{sec:autoencoder} describes how the auto-encoder was set up and trained.

\subsubsection{Diffusion Maps}\label{sec:dmaps}
The Diffusion Maps algorithm by Coifman and Lafon~\cite{coifman-2006} follows three main steps:
\begin{enumerate}
\item Construct the kernel matrix $K_{i,j}=\exp(-d(x_i,x_j)^2/\epsilon)$.
\item Normalize the matrix to mitigate effects of data density.
\item Compute eigenvectors to approximate eigenfunctions of $\Delta$ evaluated on the data.
\end{enumerate}
These steps are expanded upon in algorithm~\ref{alg:diffusion maps}. We use the implementation from \cite{lehmberg-2020}.
\begin{algorithm}[ht!]
 \caption{\label{alg:diffusion maps} Diffusion maps algorithm from~\cite{coifman-2006}, adapted from~\cite{berry-2013}, implemented in \cite{lehmberg-2020}.
 }

 \SetAlgoLined
 \textbf{Input:}
 $\ndim\in\mathbb{N}$: Dimension of points in the input.
 $\embeddingDimension\in\mathbb{N}$: Number of eigenvectors to compute.
 $X\in\R^{\numpoints\times\ndim}$: input data set, with $\numpoints$ points and $\ndim$ dimensions per point.
 \\
 \begin{enumerate}
     \item Compute kernel matrix $K\in\mathbb{R}^{\numpoints\times \numpoints}$ with $K_{ij}=\exp(-\|X_i-X_j\|^2/\epsilon)$ and kernel bandwidth  $\epsilon$ through 5\% of the median of squared distances between all points. Note that sparse computations are also possible in the software~\cite{lehmberg-2020}.
     \item Normalize kernel matrix to make it invariant to sampling density:
     \begin{enumerate}
         \item Form the diagonal normalization matrix $P_{ii}=\sum_{j=1}^N K_{ij}$.
         \item Form the matrix $W = P^{-1} K P^{-1}$.
          \item Form the diagonal normalization matrix $Q_{ii} =\sum_{j=1}^N W_{ij}$.
         \item Form the matrix $\hat{T}=Q^{-1/2} W Q^{-1/2}$.
     \end{enumerate}
     \item Solve the eigensystem and construct eigenvectors:
     \begin{enumerate}
         \item Find the $\embeddingDimension$ largest eigenvalues $a_p$ associated to non-harmonic eigenvectors $v_p$ of  $\hat{T}$. See~\cite{dsilva-2018} for a discussion and method to remove harmonic eigenvectors.
         \item Compute the eigenvalues of $\hat{T}^{1/\epsilon}$ by $\lambda_p=a_p^{1/(2\epsilon)}$.
         \item Compute the eigenvectors $\phi_p$ of the matrix $T = Q^{-1}W$ through $\phi_p=Q^{-1/2}v_p$.
     \end{enumerate}
 \end{enumerate}
 \textbf{Output:}
 $\embeddingDimension$ eigenvectors $\phi_p$, each with $\numpoints$ entries, and with corresponding eigenvalues $\lambda_p$.
\end{algorithm}

\subsubsection{SDE-informed auto-encoder}\label{sec:autoencoder}

The SDE-informed auto-encoder has a single encoder network (from observations to latent space) and two additional networks: one decoder from latent space back to the observations, and one network that defines drift and diffusivity functions \textit{on the latent space}. Note that the drift and diffusivity we find can still be considered functions of the original  observations, because $\drift(\boldsymbol{\X})\approx\driftNN(\text{encoder}(\boldsymbol{\X}))$ and $\std(\boldsymbol{\X})\approx \stdNN(\text{encoder}(\boldsymbol{\X}))$.
In the computational example that lead to \figref{fig:damp_vs_autoencoder_latent}, we used an encoder architecture of $6|25|25|25|2$, a decoder architecture $2|25|25|25|6$, and an SDE approximation network with diagonal diffusivity matrix and architecture $2|25|25|25|6$, all with \texttt{tf.nn.selu} activations.
The loss function for training the auto-encoder is the following, where $\lossfunction_{\text{Euler-Maruyama}}$ is defined in the paper (\eqnref{eq:lossfunction EM}):
\begin{equation}
    \lossfunction(\weights|\trainingdata)=\log\left(\frac{1}{B}\sum_{\idx=1}^B\left(\text{decoder}\left(\text{encoder}\left(x_0^{(\idx)}\right)\right)-x_0^{(\idx)}\right)^2\right)+\lossfunction_{\text{Euler-Maruyama}}(\weights|D),
\end{equation}
where $B$ is the size of the current mini-batch (here: $B=32$).
The Python class implementing this auto-encoder loss is called \texttt{AEModel}, and can be found in the file\\ \texttt{sde/sde\_learning\_network.py} of the repository.

\section{Additional proofs}

The logarithm of the probability density $p$ of a multivariate normal distribution in $\ndim$ dimensions is
\begin{equation}\label{eq:nd EM}
\log\left(p(\X)\right)=-\frac{\ndim}{2}\log(2\pi)-\frac{1}{2}\log\left|\det\Sigma\right|-\frac{1}{2}\left(\X-\mu\right)^T\Sigma^{-1}\left(\X-\mu\right).
\end{equation}

\begin{proof}
This follows directly from 
\begin{equation*}
p(\X)=\frac{1}{\sqrt{(2\pi)^\ndim |\det\Sigma|}}\exp\left[-\frac{1}{2}\left(\X-\mu\right)^T\Sigma^{-1}\left(\X-\mu\right)\right].
\end{equation*}
\end{proof}

\end{document}